\documentclass[preprint,showkeys]{revtex4}
\usepackage{amssymb,amsmath}
\usepackage{epsfig}
\font\bba=msbm10 % scaled 1080
\font\bbb=msbm8 %scaled 1080
\font\bbc=msbm6 %scaled 1080
\newfam\bbfam
\textfont\bbfam=\bba
\scriptfont\bbfam=\bbb
\scriptscriptfont\bbfam=\bbc
\def\bb{\fam\bbfam\bba}
\def\R{{\bb R}}

\begin{document}
\title{A new dipolar  potential  for  numerical simulations 
of polar fluids on the $4\mathrm{D}$ hypersphere.}
\author{Jean-Michel Caillol}
\email{Jean-Michel.Caillol@th.u-psud.fr}
\affiliation{Univ. Paris-Sud, CNRS, LPT, UMR 8627, Orsay, F-91405, France}    
\author{Martin Trulsson}
\email{Martin.Trulsson@u-psud.fr}  
\affiliation{Univ. Paris-Sud, CNRS, LPTMS, UMR 8626, Orsay, F-91405, France}                                              %
\date{\today}        
\begin{abstract}
We present a new method for Monte Carlo or Molecular Dynamics numerical
simulations of three dimensional polar fluids.
The simulation cell is defined to be  the surface of the northern hemisphere
of a four-dimensional (hyper)sphere. The point dipoles
are constrained to remain tangent to the sphere and their interactions are derived 
from the basic laws of electrostatics in this geometry. The dipole-dipole potential
has two singularities which correspond to the following boundary conditions : when
a dipole leaves the northern hemisphere at some point
of the equator, it reappears at the antipodal point 
bearing the same dipole moment.
We derive all the formal expressions needed to obtain the thermodynamic and structural
properties of a polar liquid at thermal equilibrium in actual numerical simulation. 
We notably establish the expression of the static dielectric constant
of the fluid as well as the behavior of the pair correlation at large distances. We report 
and discuss  the results
of extensive numerical Monte Carlo simulations for two reference states of a fluid of dipolar hard spheres
and compare these results with previous methods with a special emphasis
on finite size effects. 
\end{abstract}

\maketitle                                                                                
%%%%%%%%%%%%%%%%%%%%%%%%%%%%%%%%%%%%%%%%%%
%%%%%%%%%%%%%%%%%%%%%%%%%%%%%%%%%%%%%%%%%%
\section{Introduction}
\label{intro}
Numerical simulation of Coulomb fluids -by this terminology we mean fluids
made of charged or (and) polar molecules- need special precaution because of the
long range of electrostatics interactions. Various technical solutions to this
problem have been proposed. 
The most common one is to consider a cubic simulation cell   with periodic
boundary conditions in conjunction with Ewald summation techniques~\cite{deLeeuw,JJ}.
An alternative consists in confining particles at the surface $\mathcal{S}_3$ of a 
four-dimensional  ($4 \mathrm{D}$) 
sphere - a hypersphere for short~\cite{JJ,Caillol_1,Caillol_2_a,Caillol_2_b,Trulsson}. 
The $3 \mathrm{D}$
non-Euclidian space $\mathcal{S}_3$, albeit finite, is homogeneous and isotropic, in the sense
that it is invariant under the group $\mathcal{O}(4)$ of the $4 \mathrm{D}$
rotations; it is thus well suited for the simulation of liquids. Moreover, electrostatics
can easily be developed in $\mathcal{S}_3$ and, in particular, the Green function of Laplace equation 
can be computed analytically and it  has
a very simple expression, tailor-made  for numerical evaluations.

The present paper is devoted to dipolar fluids and we propose a new dipole-dipole potential 
in $\mathcal{S}_3$ with some advantages over  the versions considered in previous
studies~\cite{Caillol_1,Caillol_2_a,Caillol_2_b,Trulsson}.

 A brief reminder on the electrostatics in
 $\mathcal{S}_3$ should be useful  for a better understanding of these issues.
We know from Landau~\cite{Landau} that, in a finite space such as  $\mathcal{S}_3$, the total
electric charge must be equal to zero. Therefore, the building brick of electrostatics cannot
be a single point charge as we are used to in the ordinary Euclidian space $\mathbf{E}_3$.
A first possibility is to consider rather a pseudo-charge, a neologism denoting the 
association of a point charge and a uniform neutralizing background of opposite charge.
It turns out that  the electric
potential and field of a pseudo-charge can be computed analytically. Various models of
statistical mechanics involving electric charges  can therefore  be easily simulated  in  $\mathcal{S}_3$.  For instance, the one component
plasma (OCP) -\textit{i.e.} an assembly of point charges of the same sign immersed in a uniform neutralizing continuum-
may be seen as an assembly of $N$ identical pseudo-charges, the individual neutralizing back-grounds of the pseudo-charges
adding up to constitute the total neutralizing bath of the model.  High precision
Monte Carlo (MC) simulations of the thermodynamic and structural properties of the OCP have been obtained
by MC simulations of a collection of pseudo-charges  living in~$\mathcal{S}_3$~\cite{Caillol_Gilles}. 
Of course multipolar interactions are easily
derived from these basic Coulomb interactions and more complex Coulomb fluids such as
polar fluids or electrolytes can be and have  actually been
simulated before in~$\mathcal{S}_3$ , see \textit{e.g.}~\cite{Caillol_1,Caillol_2_a,Caillol_2_b,Trulsson}.

In an alternative construction of electrostatics, proposed in Ref.~\cite{Caillol_3}, the " building brick" is composed of 
a bi-charge, \textit{i.e.} 
a dumbell made  of  two antipodal charges of opposite signs $+q$ and $-q$. The potential of a bi-charge 
is obtained as a solution of Laplace-Beltrami equation  in~$\mathcal{S}_3$. It has two singularities, one at the north pole, the 
other at the south pole. 
A system of dumbells living on the whole sphere $\mathcal{S}_3$   is equivalent to a mixture of charges $+q$ and $-q$
leaving on the northern hemisphere $\mathcal{S}_3^+$. We then have the peculiar  boundary conditions : 
when the positive charge of the dumbell
leaves the northern hemisphere $\mathcal{S}_3^+$ at some point $M$
of the equator, the negative charge of the dumbell reappears at the antipodal point  $\overline{M}$ 
($\overrightarrow{O\overline{M}}=- \overrightarrow{OM}$,  $O$ center of the sphere), 
Some models with special symmetries can be
considered as made of bi-charges. For instance the restricted primitive model (RPM)
of electrolytes,  \textit{i.e.} an equimolar mixture of anions and cations of the same valence can be represented
by a simple  fluid of identical bi-charges of $\mathcal{S}_3$ (provided admittedly  that the anions and cations have the same diameter).
The extensive MC simulations of the Orsay group  on the critical point of the RPM have all 
been done in this geometry~\cite{Orsay_3}.
In the present work, bi-dipoles are built from bi-charges and used to perform actual MC simulations of
dipolar hard spheres (DHS). A fluid of bi-dipoles living on the whole surface of a hypersphere is clearly equivalent
to  a fluid of ordinary mono-dipoles living on the northern hemisphere of $\mathcal{S}_3$. When a dipole leaves the
hemisphere at some point $M$ of the equator it reenters the hemisphere at the antipodal
point $\overline{M}$, bearing the same dipolar vector. 

Our paper is organized as follows. In Sec.~\eqref{geometry} we summarize the main mathematical tools needed
in the remainder of the article. We are then well equipped to  build the electrostatics in space  $\mathcal{S}_3$
 in Sec.~\eqref{electrostatics}; starting
from Poisson's equation we  obtain the potentials and fields of  bi-charges and, by differentiation that
of bi-dipoles.  We then specialize our purpose in Sec.~\eqref{Polar_fluid} to the DHS model  in  $\mathcal{S}_3$ and derive all
formal expressions needed in MC simulations. In particular we obtain a family of formula relating the dielectric constant
to the polarization fluctuations. We also obtain the asymptotic behavior of the pair correlation function at thermal equilibrium.
In Sec.~\eqref{Simulations} we present extensive MC simulations of a DHS fluid. The models of
mono and bi-dipoles in $\mathcal{S}_3$ are compared with the the usual DHS fluid in cubico-periodical geometry.
Finite size effects on thermodynamical properties, the dielectric constant and the pair correlation
functions are studied in great detail for two reference  thermodynamic states. We conclude  in Sec.~\eqref{Conclusion}.

%%%%%%%%%%%%%%%%%%%%%%%%%%%%%%%%%%%%%%%%%%
%%%%%%%%%%%%%%%%%%%%%%%%%%%%%%%%%%%%%%%%%%
\section{Points, vectors, tensors and functions on the Hypersphere}
\label{geometry}
%%%%%%%%%%%%%%%%%%%%%%%%%%%%%%%%%%%%%%%%%%
%%%%%%%%%%%%%%%%%%%%%%%%%%%%%%%%%%%%%%%%%%
\subsection{Points and Geodesics}
\label{geodesic}
%%%%%%%%%%%%%%%%%%%%%%%%%%%%%%%%%%%%%%%%%%
%%%%%%%%%%%%%%%%%%%%%%%%%%%%%%%%%%%%%%%%%%
The simplest and most fruitful point of view is to consider the hypersphere $\mathcal{S}_3(O,R)$ of center $\mathrm{O}$ 
and radius $R$ as a  trivial generalization  of the  sphere  $\mathcal{S}_2(O,R)$ of the usual $\mathrm{3D}$ geometry.
Mathematically, it is  a compact manifold  of the $\mathrm{4D}$ Euclidian space $\mathrm{E}_4$ (to be identified  with  $\R^4$), defined as
 the subset of
points $\boldsymbol{\mathrm{OM}} = R\; (z_1, z_2,z_3,z_4)^T$   which satisfy to the constraint  \mbox{$  \mathrm{z_1^2 + z_1^2 + z_3^2 + z_4^2 = 1}$}.
When we have in mind 
the hypersphere of unit radius we adopt the uncluttered notation $  \mathcal{S}_3 \equiv   \mathcal{S}_3(O,R=1)$.
Elementary geometric constructs, valid for the sphere  $\mathcal{S}_2(\mathrm{O,R})$, can easily be extended to the $4 \mathrm{D}$ 
case~\cite{Caillol_2D,Caillol_1,Caillol_2_a} and
replace more sophisticated mathematical tools used to deal with general Riemannian manifolds.

In  $\mathcal{S}_3(O,R)$ the distance $r_{12}$ between two points $M_1$ ands $M_2$   is defined as the length
of the shortest path in the space $\mathcal{S}_3(O,R)$, \textit{i.e.} the geodesic $M_1 M_2$, 
linking these two points;
it is a bit  of the unique circle of center $\mathrm{O}$ and radius $R$  which passes
through the two points. One easily finds that
 \begin{equation}
  r_{12} /R =\psi_{12} = \cos^{-1} \left( \mathbf{z}_1 \cdot \mathbf{z}_2 \right) \; ,
 \end{equation}
where $ \mathbf{z}_i= \boldsymbol{\mathrm{OM}}_i/ R, \; i=1,2$ and $0 \leq \psi_{12} \leq \pi$.  
We denote by $\mathbf{t}_{12}(\mathrm{M_1)}$
and  $\mathbf{t}_{12}(\mathrm{M_2})$ the two unit vectors tangent to the geodesic $\mathrm{M_1}\mathrm{M_2}$,
respectively at points  $\mathrm{M_1}$ and $\mathrm{M_2}$. By convention, the arrows of the vectors point from
$M_1$ towards $M_2$. One has~\cite{Caillol_2D}

\begin{subequations}
\begin{eqnarray}
\mathbf{t}_{12}(M_1) &=&+ \frac{\mathbf{z}_2}{\sin{\psi_{12}}}-\mathbf{z}_1\cot\psi_{12} \; , \\
\mathbf{t}_{12}(M_2) &= &-\frac{\mathbf{z}_1}{\sin{\psi_{12}}}+\mathbf{z}_2\cot\psi_{12} \label{zombi}\; .
\end{eqnarray} 
\end{subequations}
Note that both vectors $\mathbf{t}_{12}(M_1)$ and $\mathbf{t}_{12}(M_2)$ are undefined for 
$\psi_{12}=0$ or $\psi_{12}=\pi$.  In the latter case  $ M_1$ and $ M_2$ are two
antipodal points and there is an infinity of geodesics, all of length $R\pi$, connecting the two points.
Henceforth we shall note  $ \overline{M}_1 $  the
antipodal point.
%%%%%%%%%%%%%%%%%%%%%%%%%%%%%%%%%%%%%%%%%%
%%%%%%%%%%%%%%%%%%%%%%%%%%%%%%%%%%%%%%%%%%
\subsection{Spherical coordinates}
\label{coord}
%%%%%%%%%%%%%%%%%%%%%%%%%%%%%%%%%%%%%%%%%%
%%%%%%%%%%%%%%%%%%%%%%%%%%%%%%%%%%%%%%%%%%
The generic unit vector $\mathbf{z}= \boldsymbol{\mathrm{OM}}/R $ of $\mathcal{S}_3$ can be  conveniently
 written in spherical coordinates as $\mathbf{z}=(\sin w  \sin v \cos u, \sin w  \sin v \sin u, \sin w  \cos v, \cos w) ^T  $ with
$0 \leq w, v  \leq  \pi$ and $0 \leq u < 2 \pi$. 
The angle $w$ determines the distance $Rw$  of point $M$ from the north pole $N$ of 
the sphere  $\mathcal{S}_3(O,R)$.
\textit{i.e.} the length of the geodesy $\mathrm{NM}$ \cite{Atkinson}. The differential vector $d \mathbf{z}$ of point  $\mathbf{z}$ of  $\mathcal{S}_3$
is easily found to be
\begin{equation}
\label{dz}
d \mathbf{z} =  dw \,  \mathbf{e}_w  + \sin w dv \, \mathbf{e}_v +  \sin w \sin v \, du \, \mathbf{e}_u \; , 
\end{equation}
with
\begin{subequations}
\begin{align}
\mathbf{e}_w &\equiv \partial \mathbf{z}/\partial w =(\cos w  \sin v \cos u, \cos w  \sin v \sin u, \cos w  \cos v,- \sin w)^T \; , \\
\mathbf{e}_v &\equiv ( \partial \mathbf{z}/\partial v)  / \sin w =( \cos v \cos u,  \cos v \sin u,  -\sin v, 0)^T \; , \\
\mathbf{e}_u &\equiv( \partial \mathbf{z}/\partial u) / (\sin w \sin v) =( - \sin u,  \cos u,  0 , 0)^T \; .
\end{align}
\end{subequations}
The 3 orthonormal vectors $(\mathbf{e}_u , \mathbf{e}_v, \mathbf{e}_w) $ constitute the '' local basis'' of   $\mathcal{S}_3$
in spherical coordinates.  This basis spans the $\mathrm{3D}$ Euclidian space $\mathcal{T}_{3}(M)$, 
tangent to the hypersphere at point  $M$.
To make some contact with the material of section~\eqref{geodesic} we note that 
$\mathbf{e}_w(\mathbf{z}) =\mathbf{t}_\mathrm{{NM}} \mathrm{(M)}$ is the unit vector, tangent at the geodesic  $\mathrm{NM}$
 at point $M$.
Moreover one checks readily that it satisfies  identity~\eqref{zombi}.

It also follows from Eq.~\eqref{dz} that  the infinitesimal length element of   $\mathcal{S}_3(O,R)$ is
$ds^2=\sin^2 w \sin^2 u \;  du^2 + \sin^2 w   \;  dv^2 +  \;  dw ^2 $ and that the infinitesimal volume element 
takes the simple form  $d \tau = R^3 d \Omega =  R^3 \sin^2 w  \sin v  \,du \, dv \, dw $, so that the  total
volume of space $\mathcal{S}_3(O,R)$ is  \mbox{$V_{\mathrm{Tot.}}= \int d\tau = 4 \pi^2 R^3$}.

It is in place to
define the unit dyadic tensor $\mathbf{U}_{\mathcal{S}_3}(\mathbf{z})= \mathbf{e}_u \mathbf{e}_u
+ \mathbf{e}_v \mathbf{e}_v +  \mathbf{e}_w \mathbf{e}_w $ of the tangent Euclidian space  $\mathcal{T}_{3}(\mathbf{z})$;
note
that the unit dyadic tensor of Euclidian space $\mathrm{E}_4$ is clearly given by $\mathbf{U}_{\R^4}=
\mathbf{U}_{\mathcal{S}_3}(\mathbf{z})
+ \mathbf{z} \mathbf{z} $. These admittedly old-fashioned objects however allow
an easy definition of the gradient in  $\mathcal{S}_3$,  or  first differential  Beltrami operator, 
as
\begin{equation*}
 \nabla_{\mathcal{S}_3}=    \mathbf{U}_{\mathcal{S}_3}(\mathbf{z})  \cdot \nabla_{\R^4} \; ,
\end{equation*}
where 
$\nabla_{\R^4}$ is the usual Euclidian gradient operator of $\R^4$ and 
the dot in the r.h.s. denotes the $\mathrm{4D}$ tensorial contraction. 
Note that the gradient in the hypersphere  $\mathcal{S}_3(O,R)$ of radius $R \neq 1$ is
of course defined as $\nabla_{\mathcal{S}_3  \mathrm{(O,R)}}=  \nabla_{\mathcal{S}_3}/R$.

The Laplace-Beltrami operator
(or  second differential  Beltrami operator) will be similarly defined as the restriction of the $\mathrm{4D}$
Laplacian $\Delta_{\R^4}$ to the unit sphere. One has~\cite{Atkinson}
\begin{equation*}
 \Delta_{\mathcal{S}_3(\mathrm{0,R})} \equiv  \Delta_{\mathcal{S}_3}/ R^2=  \Delta_{\R^4} -\frac{\partial^2}
{ \partial R^2} -\frac{3}{R} \frac{\partial }{\partial R }  \; .
\end{equation*}
Many theorems of vectorial analysis involving Betrami operators find their counterpart in the space $\mathcal{S}_3$. This is notably
the case of the Green-Beltrami theorem which extends the well known Green's first identity~\cite{Jackson} and 
is of an overwhelming importance to build the electrostatics in  $\mathcal{S}_3$.
It reads~\cite{Atkinson}:
\begin{equation}
 \label{Beltrami}
 \int_{\mathcal{S}_3} d\Omega \;  \nabla_{\mathcal{S}_3}f \cdot \nabla_{\mathcal{S}_3}g = -  \int_{\mathcal{S}_3} d\Omega \; f  \Delta_{\mathcal{S}_3}g \; ,
\end{equation}
where $f(\mathbf{z})$ and  $g(\mathbf{z})$ are functions defined on the unit sphere $\mathcal{S}_3$.
The missing proof of theorem~\eqref{Beltrami} (as well as the proofs of many other statements given in
the sequel) is not so difficult and can be found in the recent textbook by Atkinson and Han~\cite{Atkinson}.

%%%%%%%%%%%%%%%%%%%%%%%%%%%%%%%%%%%%%%%%%%
%%%%%%%%%%%%%%%%%%%%%%%%%%%%%%%%%%%%%%%%%%
\subsection{Functions defined on  $\mathcal{S}_3(O,R)$}
\label{functions}
%%%%%%%%%%%%%%%%%%%%%%%%%%%%%%%%%%%%%%%%%%
%%%%%%%%%%%%%%%%%%%%%%%%%%%%%%%%%%%%%%%%%%
The eigenfunctions of the Beltrami-Laplace operator $\Delta_{\mathcal{S}_3}$ are
 the $\mathrm{4D}$ spherical harmonics $Y_{L,\boldsymbol{\alpha}}(\mathbf{z})$
with eigenvalues $-L(L+2)$, $L=0, 1, \ldots$ being   a positive integer; \textit{i.e.} one has
\begin{equation}
 \label{vp}
\Delta_{\mathcal{S}_3} Y_{L,\boldsymbol{\alpha}} = -L(L+2) Y_{L,\boldsymbol{\alpha}} \; .
\end{equation}
The degeneracy of the eigenvalue labelled by $L$ is $(L+1)^2$ 
and the second  "quantum" number
$\boldsymbol{\alpha}$ accounts for this degeneracy. Its precise algebraic structure depends of the representation
of the spherical harmonics. Quite generally, in a space $\mathbf{E}_3$ of arbitrary dimension $\mathrm{D=4}$, the spherical harmonics
 $Y_{L,\boldsymbol{\alpha}}(\mathbf{z})$
is a harmonic and  homogeneous polynomial of  $\mathrm{D}$ variables  and 
degree $\mathrm{L}$ restricted to the unit sphere
$\mathcal{S}_{(D-1)}$ (in this paper  $\mathrm{D=4}$)~\cite{Atkinson}. This has   the interesting consequence  that
$Y_{L,\boldsymbol{\alpha}}(-\mathbf{z})=(-1)^LY_{L,\boldsymbol{\alpha}}(\mathbf{z})$.
Explicit expressions of the $Y_{L,\boldsymbol{\alpha}}(\mathbf{z})$
in spherical coordinates will be found in Refs.~\cite{Avery,Vilenkin,Higuchi} but are of little use in these lines.
More important is the fact that the $\mathrm{4D}$ spherical harmonics 
$Y_{L,\boldsymbol{\alpha}}(\mathbf{z})$ constitute a complete basis set to expand functions $f(\mathbf{z})$ defined
on the unit hypersphere  $\mathcal{S}_3$. Orthogonality and completeness relations take the following form:
\begin{subequations}
\begin{eqnarray}
 \int_{\mathcal{S}_3}  d\, \Omega \;  Y^{*}_{L,\boldsymbol{\alpha}}(\mathbf{z}) Y_{L^{'},\boldsymbol{\alpha}^{'}}(\mathbf{z}) &=& \delta_{L L^{'}}
\delta_{\boldsymbol{\alpha} \boldsymbol{\alpha}^{'}}  \label{ortho}\, , \\ 
 \sum_{L,\boldsymbol{\alpha}} 
Y^*_{L,\boldsymbol{\alpha}}(\mathbf{z}) Y_{L,\boldsymbol{\alpha}}(\mathbf{z})
&=&\delta_{\mathcal{S}_3}(\mathbf{z},\mathbf{z}^{'}) \, ,
\end{eqnarray}
\end{subequations}
where the delta function $\delta_{\mathcal{S}_3}(\mathbf{z},\mathbf{z}^{'}) \equiv \delta (1-\mathbf{z} \cdot \mathbf{z}^{'})$~\cite{Atkinson}
has the usual  convolution property 
\begin{equation}
  \int_{\mathcal{S}_3}  d\, \Omega^{'}  \, f(\mathbf{z}^{'}) \delta_{\mathcal{S}_3}(\mathbf{z},\mathbf{z}^{'}) = f(\mathbf{z}) \; .
\end{equation}
The delta function on the sphere $\mathcal{S}_3(O,R)$ will be conveniently denoted by
\begin{equation}
 \delta (M,M^{'})=\delta_{\mathcal{S}_3}(\mathbf{z},\mathbf{z}^{'}) / R^3.
\end{equation}

Moreover, as in $D=3$, the $\mathrm{4D}$ harmonics satisfy a so-called addition theorem which reads: 
\begin{subequations}
 \begin{eqnarray}
 \sum_{\boldsymbol{\alpha}} 
Y^*_{L, \boldsymbol{\alpha}}(\mathbf{z}) Y_{L,\boldsymbol{\alpha}}(\mathbf{z}^{'})&=& P_L(\mathbf{z} \cdot \mathbf{z}^{'}) \; , \\
P_L(\cos(\psi))&=  & \frac{L+1}{2 \pi^2} \frac{\sin( (L+1) \psi)}{\sin \psi}\; ,
 \end{eqnarray}
\end{subequations}
where  the Tchebycheff polynomials of the second kind $P_L(\cos(\psi))$ play, in $D=4$,
 the role devoted to the Legendre polynomials in $D=3$.
%%%%%%%%%%%%%%%%%%%%%%%%%%%%%%%%%%%%%%%%%%
%%%%%%%%%%%%%%%%%%%%%%%%%%%%%%%%%%%%%%%%%%
\subsection{Vectors and vector fields of $\mathcal{S}_3(O,R)$}
\label{fields}
By convention, a vector $\boldsymbol{\mu}$ of $\mathcal{S}_3(O,R)$ at point $M$  should
be an ordinary vector of  the   $3\mathrm{D}$ 
Euclidian space $\mathcal{T}(M)$, tangent to the hypersphere at point 
 $M$.  Taking the scalar product of
two vectors  $\boldsymbol{\mu}_1$ and $\boldsymbol{\mu}_2$ located at two distinct points,
 $\mathrm{M_1}$ and $\mathrm{M_2}$ of $\mathcal{S}_3(O,R)$ needs some precaution. It first requires 
to perform a parallel transport of vector   $\boldsymbol{\mu}_1$ from 
 $\mathrm{M_1}$ to $\mathrm{M_2}$ along the geodesic $\mathrm{M_1}\mathrm{M_2}$ and then to take a $3\mathrm{D}$
scalar product in space  $\mathcal{T}(M_2)$.
Thus~\cite{Caillol_2D,Caillol_1} 
\begin{equation}
\label{scal}
 \left< \boldsymbol{\mu}_1 , \boldsymbol{\mu}_2\right > = \tau_{12} \boldsymbol{\mu}_1 \cdot \boldsymbol{\mu}_2 \; ,
\end{equation}
where, in the r.h.s. the dot denotes the usual scalar product of the Euclidian space 
\mbox{$ \mathcal{T}(M_2) \subset  \mathrm{E}_4$.}
Vector  $\tau_{12} \boldsymbol{\mu}_1 $  results from a  transport of $\boldsymbol{\mu}_1$ 
from the space  $\mathcal{T}(M_1)$ to  the space  $\mathcal{T}(M_2)$ along the geodesic
 $\mathrm{M_1}\mathrm{M_2}$, keeping its angle with the tangent to the geodesic constant. Explicitely
one has:
\begin{equation}
 \label{transport}
 \tau_{12} \boldsymbol{\mu}_1 =\boldsymbol{\mu}_1 - \frac{ \boldsymbol{\mu}_1 \cdot \mathbf{z}_2}{1+\cos \psi_{12}} \;  
\left(  \mathbf{z}_1+  \mathbf{z}_2  \right)
\end{equation}
One checks the following geometrical properties
\begin{align}
  \tau_{12} \boldsymbol{\mu}_1 \cdot \mathbf{z}_2 &= 0 \; , \nonumber \\
 \tau_{12} \mathbf{t}_{12} (1 ) &= \mathbf{t}_{12} (2)  \; ,  \nonumber \\
  \tau_{12}   \tau_{21} \boldsymbol{\mu}_1  &= \boldsymbol{\mu}_1 \; .
\end{align}
Finally, by taking into account Eq.~\eqref{transport}, the scalar product~\eqref{scal} may be rewritten more explicitely as
\begin{equation}
\label{scal_bis}
  \left< \boldsymbol{\mu}_1 , \boldsymbol{\mu}_2\right > = \boldsymbol{\mu}_1 \cdot \boldsymbol{\mu}_1
- \frac{(  \boldsymbol{\mu}_1 \cdot \mathbf{z}_2) \cdot (  \boldsymbol{\mu}_2 \cdot \mathbf{z}_1)    }{ 1+\cos \psi_{12} } \; .
\end{equation}

Besides the scalar fields of section~\eqref{functions} one also  needs to consider vector fields.
An example will be a field of gradients. Let $f(\mathbf{z}_1,\mathbf{z}_2) $ be
some scalar field of \textit{two} variables defined on
the unit sphere $\mathcal{S}_3$.
 We suppose that the two-point function $f(\mathbf{z}_1,\mathbf{z}_2)$
is invariant under the rotations of
the Euclidian space $\mathrm{E}_4$ which leave the center $\mathrm{O}$ invariant
(\textit{i.e.} the rotations of the orthogonal group $\mathcal{O}(4)$).
Therefore  $f(\mathbf{z}_1,\mathbf{z}_2) \equiv \widetilde{f}(\psi_{12})$ depends solely  on the geodesic length $\psi_{12}$. 
Taking the gradients of
$f(\mathbf{z}_1,\mathbf{z}_1) $
at points $\mathbf{z}_1$ or $\mathbf{z}_2$ defines two gradient fields, obviously given by:
 \begin{align}
\label{grad}
   \nabla_{\mathcal{S}_{3, 1}} f(\mathbf{z}_1,\mathbf{z}_2) = - &\frac{\partial  \widetilde{f}(\psi_{12})}{\partial \psi_{12}} \;\mathbf{t}_{12}(\mathbf{z}_1) \nonumber \; , \\
 \nabla_{\mathcal{S}_{3, 2}}  f(\mathbf{z}_1,\mathbf{z}_2)  =+&  \frac{\partial  \widetilde{f}(\psi_{12})}{\partial \psi_{12}} \;\mathbf{t}_{12}(\mathbf{z}_2)  \; .
 \end{align}

%%%%%%%%%%%%%%%%%%%%%%%%%%%%%%%%%%%%%%%%%%
%%%%%%%%%%%%%%%%%%%%%%%%%%%%%%%%%%%%%%%%%%
\section{Elementary Electrostatics of  $\mathcal{S}_3(O,R)$}
\label{electrostatics}
%%%%%%%%%%%%%%%%%%%%%%%%%%%%%%%%%%%%%%%%%%
%%%%%%%%%%%%%%%%%%%%%%%%%%%%%%%%%%%%%%%%%%
\subsection{Poisson Equation}
\label{Poisson}
Given a charge distribution $\rho_{\mathcal{S}_3}(\mathbf{z})$ of  $\mathcal{S}_3$,
the electric potential  $V_{\mathcal{S}_3}(\mathbf{z})$ is  defined to be
the solution of  Poisson's equation
\begin{equation}
\label{Poisson_eq}
  \Delta_{\mathcal{S}_3} V_{\mathcal{S}_3} =
-4 \pi   \rho_{\mathcal{S}_3} \;, 
\end{equation}
where the operator entering the r.h.s. of the equation is the Laplace-Beltrami operator of section~\eqref{coord}.
We first note that making $f=1$ and $g=V_{\mathcal{S}_3}$ in equation~\eqref{Beltrami} implies
that the integral of  $\Delta_{\mathcal{S}_3} V_{\mathcal{S}_3}$ over the whole hypersphere is zero.
It follows, as a consequence of Poisson's equation~\eqref{Poisson_eq},  that the total charge of
the space must vanish. As already pointed out in the introduction,
the potential of a single point charge is not defined in $\mathcal{S}_3$. Elementary objects need be neutral.
In this paper we consider electrostatics based on bi-charges~\cite{Caillol_3}.

%%%%%%%%%%%%%%%%%%%%%%%%%%%%%%%%%%%%%%%%%%
%%%%%%%%%%%%%%%%%%%%%%%%%%%%%%%%%%%%%%%%%%
\subsection{Bi-charges and bi-dipoles}
\label{bidip}
We first consider a bi-charge $q$ at point $M_0$ of $\mathcal{S}_3(O,R)$,
\textit{i.e.} a dumbell made  of a point charge $+q$ at point $M_0$ and a point charge $-q$ at 
the antipodal point $\overline{M}_0$,
 with $\boldsymbol{\mathrm{O}\overline{M}}_0 = - \boldsymbol{\mathrm{O}M}_0$. It can be denoted
as  $(M_0, q ) \cup (\overline{M}_0, -q )$.
The potential $V_{M_0}(M)$ created by $(M_0, q ) \cup (\overline{M}_0, -q )$ at a point 
$M$ of $\mathcal{S}_3(O,R)$ satisfies to Poisson equation :
\begin{equation}
\label{Poisson_bi}
  \Delta_{\mathcal{S}_3(\mathrm{0,R})} V_{M_0}(M) =
-\frac{4 \pi  q}{R^3} \left(   \delta_{\mathcal{S}_3}(\mathbf{z},\mathbf{z}_{0})    
                         -     \delta_{\mathcal{S}_3}(\mathbf{z},\overline{\mathbf{z}}_{0})         \right) \; ,
\end{equation}
with the obvious notations
 $\overline{\mathbf{z}}_{0} = - \mathbf{z}_{0}=-\boldsymbol{\mathrm{O}M}_0/R $.
Expanding both sides of \eqref{Poisson_bi} upon spherical harmonics yields~\cite{Caillol_1} :
 \begin{align}
\label{kluk}
   V_{M_0}(M) &= \frac{8 \pi}{R} \sum_{L \;, \boldsymbol{\alpha}} \,^{'}
\frac{1}{L(L+2)} Y^*_{L,\boldsymbol{\alpha}}(\mathbf{z_0}) Y_{L,\boldsymbol{\alpha}}(\mathbf{z}) \nonumber \; \\
&= \dfrac{q}{R} \cot \psi_{M_0M} \; ,
 \end{align}
where the prime  affixed to the sum in~\eqref{kluk} denotes the restriction that $L$ is an odd, positive integer.
Notice that the potential is singular for $\psi_{M_0M}=0$ and $\psi_{M_0M}= \pi$.  At a given $r=R \psi_{M_0M}$ and in 
the large $R$ limit, one recovers the Euclidian behavior $  V_{M_0}(M) \sim q/r$, and, at the antipodal  point
 $  V_{M_0}(M) \sim -q/r$ as expected.

The potential created at point $M$ by  a bi-dipole $\boldsymbol{\mu}_0$ located at point   $M_0$
is now obtained by a standard limit process :
 \begin{align}
    V_{M_0, \boldsymbol{\mu}_0}(M) &=  
  \dfrac{\boldsymbol{\mu}_0 }{R} \cdot \nabla_{\mathcal{S}_3, M_0}  V_{M_0}(M)    \; , \nonumber \\
&=\frac{\mu}{R^2}  \frac{1}{\sin^2( \psi_{M_0M})} \; \mathbf{s}_0 \cdot 
\mathbf{t}_{M_0 M} (M_0) \; ,   %% \\
%% &= \frac{\mu}{R^2}    \frac{1}{\sin^3( \psi_{M_0M})}   \;   \mathbf{s}_0 \cdot \mathbf{z}
 \end{align}
where $\mathbf{s}_0 = \boldsymbol{\mu}_0/ \mu$ is the direction of $\boldsymbol{\mu}_0$ and $ \mu$
its modulus.
It can be remarked that our bi-dipole can be seen as the dumbell
 $(M_0, \boldsymbol{\mu}_0 ) \cup (\overline{M}_0, \boldsymbol{\mu}_0 )$.
The  dipolar potential   $V_{M_0, \boldsymbol{\mu}_0}(M)$ is of course a
fundamental, non-isotropic solution of Laplace equation on the hypersphere.
Note that in the  limit  $r=R \psi_{M_0M}$ fixed, $R \to \infty$, vector
 $\mathbf{t}_{M_0 M} (M_0) \sim  \widehat{\mathbf{r}} = \overrightarrow{M_0M}/M_0M$ and one recovers
the Euclidian expression   $V_{M_0, \boldsymbol{\mu}_0}(M) \sim \boldsymbol{\mu}_0 \cdot  
  \widehat{\mathbf{r}}/r^2$ as expected.

 The electric field created by the dipole  is obtained by taking minus the gradient of 
$V_{M_0, \boldsymbol{\mu}_0}(M)$ at point $M$ with the result :
\begin{equation}
\label{field}
 \mathbf{E}_{M_0, \boldsymbol{\mu}_0}(M) = -
 \dfrac{1}{R} \cdot \nabla_{\mathcal{S}_3, M}  V_{M_0}(M)  = 4 \pi
\mathbf{G}_0 (M, M_0) \cdot \boldsymbol{\mu}_0\; ,
\end{equation}
where we have introduced the tensorial vectorial Green's function $\mathbf{G}_0 (M,
M_0) $ for which we can give two expressions :

\begin{subequations}
\label{G0}
  \begin{align}
\mathbf{G}_0 (M,M_0) =& \frac{1}{4 \pi R^3} \frac{1}{\sin^3( \psi_{M_0M})}  \,
 [ 3  \cos\left( \psi_{M_0M}\right)  \mathbf{t}_{M_0 M} (M)\mathbf{t}_{M_0 M} (M_0) - \nonumber \label{G0_a} \\
&\mathbf{U}_{\mathcal{S}_3}(\mathbf{z}) \cdot  \mathbf{U}_{\mathcal{S}_3}(\mathbf{z}_0)] 
  \; ,    \\
&=- \frac{2}{R^3} \sum_{L \;, \boldsymbol{\alpha}} \, ^{'}  \frac{1}{L(L+2)} \nabla_{\mathcal{S}_3}Y^*_{L,\boldsymbol{\alpha}}(\mathbf{z})
          \nabla_{\mathcal{S}_3}  Y_{L,\boldsymbol{\alpha}}(\mathbf{z}_0)  \; , \label{G0_b}
 \end{align}
\end{subequations}
as a short algebra will show. 

We stress that $\mathbf{G}_0 (M,M_0)$ is a $4 \mathrm{D}$ dyadic
tensor of the type $\mathbf{A}(M) \mathbf{A}(M_0)$, $\mathbf{A}(M)$ and $ \mathbf{A}(M_0)$ being
two vectors tangent to the hypersphere at the points $M$ and $M_0$, respectively. It is easy to show that
in the limit $\psi_{M_0M} \to 0$, $\mathbf{G}_0 (M,M_0)$ tends to its Euclidian limit
$\mathbf{G}_{0, \R^3} (M,M_0)=  [-\mathbf{U}_{\mathcal{S}_3}(M) + 3 \widehat{\mathbf{r}}
 \widehat{\mathbf{r}} ] /(4 \pi r^3)$, with as usual $\mathbf{r} = \overrightarrow{M_0M}$ 
and $\widehat{\mathbf{r}}=  \mathbf{r} /r$. The distribution $\mathbf{G}_{0, \R^3} (M,M_0)$ has a singularity
$-(1/3) \mathbf{U} \delta(\mathbf{r})$~\cite{Jackson,Fulton} and therefore $\mathbf{G}_0 (M,M_0)$ is singular
for $\psi_{M_0M} \to 0$, with the same singularity. It may be important to extract this singularity and to define
a non-singular Green function  $\mathbf{G}_0^{\delta} (M,M_0)$ by the relations
\begin{subequations}
 \begin{align} \label{decompo} 
 \mathbf{G}_0(M,M_0) & =   \mathbf{G}_0^{\delta} (M,M_0) + \frac{1}{3} \delta(M, M_0)
                                             \mathbf{U}_{\mathcal{S}_3}(\mathbf{z}) \; ,  & \\
\mathbf{G}_0^{\delta} (M,M_0)  &= \begin{cases}
                                                  \mathbf{G}_0(M,M_0)\; &,   \text{ for } R  \psi_{M_0M} > \delta  \; ,  \\        
                                                  0                        \;  &,    \text{ for } R  \psi_{M_0M} < \delta  \; , 
                                                     \end{cases}
 \end{align}
\end{subequations}
where $\delta$ is an arbitrary small cut-off ultimately set to zero. It must be understood that any integral
involving $\mathbf{G}_0^{\delta} $ must be calculated with $\delta \neq 0$
 and then taking the limit $\delta \to 0$.
Some useful mathematical properties of $\mathbf{G}_0 (M,M_0)$ are derived in the appendix.

We end this section by \textit{defining } the interaction of two bi-dipoles $(M_1,\boldsymbol{\mu}_1)$
and $(M_2,\boldsymbol{\mu}_2)$ as $W_{\boldsymbol{\mu}_1,  \boldsymbol{\mu}_2} \equiv - \boldsymbol{\mu}_1 \cdot
4 \pi  \mathbf{G}_0(1,2) \cdot \boldsymbol{\mu}_2 $ which gives, more explicitely and with the help of Eq.~\eqref{G0_a}

\begin{subequations}
\begin{align}
 W_{\boldsymbol{\mu}_1 ,\boldsymbol{\mu}_2} &= \frac{1}{R^3}\frac{1}{\sin^3 \psi_{12}}
\bigg(  \boldsymbol{\mu}_1  \cdot     \boldsymbol{\mu}_2  
 -    3 \cos \psi_{12} ( \mathbf{t}_{12}(1) \cdot   \boldsymbol{\mu}_1 )
  ( \mathbf{t}_{12}(2) \cdot   \boldsymbol{\mu}_2 ) \bigg)  \; ,  \\
&= \frac{1}{R^3}\frac{1}{\sin^3 \psi_{12}} \bigg(  \boldsymbol{\mu}_1  \cdot     \boldsymbol{\mu}_2 
+3 \frac{\cos \psi_{12}}{\sin^2 \psi_{12}}  (  \boldsymbol{\mu}_1 \cdot \mathbf{z}_2)
 ( \boldsymbol{\mu}_2 \cdot\mathbf{z}_1)
 \bigg) \; .  \label{dip_bi}
\end{align}
\end{subequations}
Once again one recovers the well-known Euclidian limit \mbox{
$ W_{\boldsymbol{\mu}_1 ,\boldsymbol{\mu}_2} \sim (1/r_{12}^3) [\boldsymbol{\mu}_1 \cdot \boldsymbol{\mu}_1 
-3( \boldsymbol{\mu}_1 \cdot \widehat{\mathbf{r}}_{12})  ( \boldsymbol{\mu}_2 \cdot \widehat{\mathbf{r}}_{12})   ]$}
of the dipole-dipole interaction when \mbox{$\psi_{12} \to 0$}.

%%%%%%%%%%%%%%%%%%%%%%%%%%%%%%%%%%%%%%%%%%
%%%%%%%%%%%%%%%%%%%%%%%%%%%%%%%%%%%%%%%%%%
\section{Polar fluid in  $\mathcal{S}_3(O,R)$}
\label{Polar_fluid}
%%%%%%%%%%%%%%%%%%%%%%%%%%%%%%%%%%%%%%%%%%
%%%%%%%%%%%%%%%%%%%%%%%%%%%%%%%%%%%%%%%%%%
\subsection{Two models of polar hard spheres in  $\mathcal{S}_3(O,R)$}
\label{HS}
We consider two versions of a fluid of  $N$ dipolar hard spheres in $\mathcal{S}_3(O,R)$.

%%%%%%%%%%%%%%%%%%%%%%%%%%%%%%%%%%%%%%%%%%
\subsubsection{Mono-dipoles}
\label{Mono-dipoles}
%%%%%%%%%%%%%%%%%%%%%%%%%%%%%%%%%%%%%%%%%%
The first version is that already considered in Ref.~\cite{Caillol_2_b}. The dipoles are
formed from pseudo-charges and are confined on the surface of the hypersphere. They
must be carefully distinguished from those of Sec.~\eqref{electrostatics} which are formed
from bi-charges and take the appearance of dumbells of dipoles.
In a given configuration
of point-dipoles $\boldsymbol{\mu}_i$ located at the points $OM_i=R \mathbf{z}_i$ ($i=1, \ldots,N$)
of $\mathcal{S}_3(O,R)$ the configurational energy reads
\begin{equation}
\label{ }
 U =  \frac{1}{2} \sum_{i \neq j}^N \; v_{\mathrm{HS}}^{\mathrm{mono}}(\psi_ {ij}) + \frac{1}{2}
\sum_{i \neq j}^N \;  W_{\boldsymbol{\mu}_i, \boldsymbol{\mu}_j}^{\mathrm{mono}}  \; ,
\end{equation}
where $v_{\mathrm{HS}}^{\mathrm{mono}}(\psi_ {ij}) $ is the hard-core pair potential in $\mathcal{S}_3(O,R)$
defined by
\begin{equation}
 v_{\mathrm{HS}}^{\mathrm{mono}}(\psi_ {ij}) =  \begin{cases}
                                                 \infty & \text{ if }  \sigma/R > \psi_{ij}   \; ,   \\
                                                 0&       \text{ otherwise } \; ,  
                                                \end{cases}
\end{equation}
and $ W_{\boldsymbol{\mu}_i, \boldsymbol{\mu}_j}^{\mathrm{mono}}$ is the energy of a pair of mono-dipoles.
Recall that
\begin{align}
\label{dip_mono}
 W_{\boldsymbol{\mu}_i ,\boldsymbol{\mu}_j}^{\mathrm{mono}}  =&
   \frac{1}{\pi R^3} \bigg[ \frac{2}{\sin^2 \psi_{ij}}  ( \boldsymbol{\mu}_i \cdot\mathbf{z}_j)( \boldsymbol{\mu}_j \cdot\mathbf{z}_i)  \nonumber \\
& + f(\psi_{ij}) \; \bigg( \boldsymbol{\mu}_i \cdot\boldsymbol{\mu}_j +3 \cot \psi_{ij}
\frac{( \boldsymbol{\mu}_i \cdot\mathbf{z}_j)( \boldsymbol{\mu}_j \cdot\mathbf{z}_i) }{\sin \psi_{ij}}
\bigg) \bigg] \;
\end{align}
with 
\begin{equation}
 f(\psi_{ij}) =\frac{1}{\sin \psi_{ij}} \left(\cot \psi_{ij} + \dfrac{\pi - 
\psi_{ij}}{\sin^2 \psi_{ij}} \right) \; .
\end{equation}

We want to stress that the electric potentials created by mono- and bi-dipoles are both fundamental solutions of 
Laplace-Beltrami equation in the space $\mathcal{S}_3(O,R)$. These solutions differ by their singularities at
$\psi = 0 $ (mono- and bi-dipoles) and  $\psi = \pi$ (bi-dipoles). The resulting dipole-dipole interactions
are quite different as apparent on Eqs.~\eqref{dip_bi} (bi-dipoles) and~\eqref{dip_mono} (mono-dipoles).
However it is noteworthy that both interactions indeed present the same Euclidian limit for $\psi_{ij} \to 0$.

A thermodynamic state of this model is characterized by a density $\rho^*=N\sigma^3/V$ where $V=2\pi^2R^3$
is the $3\mathrm{D}$ surface  of the hypersphere $\mathcal{S}_3(O,R)$ and a reduced inverse temperature
$\mu^*$ with  $\mu^{*2}= \mu^2/(k_B T\sigma^3) $ ($k_B $ Boltzmann constant, $T$ absolute temperature in Kelvin).
%%%%%%%%%%%%%%%%%%%%%%%%%%%%%%%%%%%%%%%%%%
\subsubsection{Bi-dipoles}
\label{bi-dipoles}
%%%%%%%%%%%%%%%%%%%%%%%%%%%%%%%%%%%%%%%%%%
The second version is that introduced in Sec.~\eqref{electrostatics}, \textit{i.e.} a fluid of bi-dipoles
confined on the surface of the hypersphere $\mathcal{S}_3(O,R)$.
 Clearly, as in the case of bi-charges (cf.~\cite{Caillol_3})
both dipoles of the dumbell must be embedded at the center of a hard sphere of diameter
$\sigma$ to avoid a collapse of the system.
In a given configuration
of $N$ bi-dipoles $\boldsymbol{\mu}_i$ located at the points $OM_i=R \mathbf{z}_i$ ( $i=1, \ldots,N$)
of $\mathcal{S}_3(O,R)$ the configurational energy reads 
\begin{equation}
\label{conf2}
 U =  \frac{1}{2} \sum_{i \neq j}^N \; v_{\mathrm{HS}}^{\mathrm{bi}}(\psi_ {ij}) + \frac{1}{2}
\sum_{i \neq j}^N \;  W_{\boldsymbol{\mu}_i, \boldsymbol{\mu}_j}^{\mathrm{bi}}  \; ,
\end{equation}
where $v_{\mathrm{HS}}^{\mathrm{mono}}(\psi_ {ij}) $ is  hard-core pair potential  defined by
\begin{equation}
 v_{\mathrm{HS}}^{\mathrm{bi}}(\psi_ {ij}) =  \begin{cases}
                                                 \infty & \text{ if }  \sigma/R > \psi_{ij}  \text{ or }    \psi_{ij}> \pi -\sigma/R  \; ,   \\
                                                 0&       \text{ otherwise }  \; ,  
                                                \end{cases}
\end{equation}
and the dipole-dipole interaction $W_{\boldsymbol{\mu}_i, \boldsymbol{\mu}_j}^{\mathrm{bi}} $ is precisely
that defined at Eq.~\eqref{dip_bi}.

The interpretation of this seemingly strange model is the following. It is easily realized that the genuine  domain
occupied by
the model is  the northern hemisphere  $\mathcal{S}_3(O,R)^+$ rather than the whole
hypersphere. When a dipole $\boldsymbol{\mu}$ quits $\mathcal{S}_3(O,R)^+$ at some point $M$ of the equator
the same $\boldsymbol{\mu}$  reenters at the antipodal point  $\overline{M}$. So bi-dipoles living
on the whole sphere are equivalent to mono-dipoles living on a single hemisphere but with boundary conditions
which ensure homogeneity and isotropy at equilibrium. Other boundary conditions ensuring homogeneity and isotropy
could be invented but yield more complicated dipolar interactions.

A thermodynamic state of this model is now characterized by a density $\rho^*=N\sigma^3/V$ where $V=\pi^2R^3$
is the $3\mathrm{D}$ surface  of the northern hemisphere  $\mathcal{S}_3(O,R)^+$ and the  reduced inverse temperature
$\mu^*$ with  $\mu^{*2}= \mu^2/(k_B T\sigma^3) $ as in Sec.~\eqref{Mono-dipoles}. 

%%%%%%%%%%%%%%%%%%%%%%%%%%%%%%%%%%%%%%%%%%
%%%%%%%%%%%%%%%%%%%%%%%%%%%%%%%%%%%%%%%%%%
\subsection{Thermodynamics and structure}
\label{Thermo}
The thermal average of the energy per particle $ u= \langle U \rangle/N $ as well as other thermodynamic quantities should
be the same for both models in a given state $(\rho^*, \mu^{*2})$, at least in the thermodynamic limit $N \to \infty$
with $\rho^*$ fixed. We have checked this point by means of extensive MC simulations in the canonical ensemble and
we postpone the discussion of these numerical experiments to Sec.~\eqref{Simulations}.

The structure at equilibrium is also of prime importance~\cite{JJ,Hansen}. In the fluid phase of a molecular liquid of linear molecules
the  equilibrium pair correlation function can be expanded on a set of rotational invariants
 $\Phi^{lmn}(1,2)$~\cite{Blum,Caillol_2_a} as
\begin{equation}
 g(1,2) =g^{000}(r_{12})  \Phi^{000}(1,2) + h^{110}(r_{12})  \Phi^{110}(1,2) +   h^{112}(r_{12})  \Phi^{112}(1,2) + \ldots
\end{equation}

with, in $\mathcal{S}_3$~\cite{Caillol_2_a},
\begin{subequations}
\label{fi}
 \begin{align}
  \Phi^{000}(1,2) & =1  \; ,\\
  \Phi^{110}(1,2) & = <\mathbf{s}_1 , \mathbf{s}_2> \; \nonumber,\\
                          &= \mathbf{s}_1 \cdot \mathbf{s}_2 - \frac{1}{1+ \cos \psi _{12} } \left(\mathbf{s}_1 \cdot  \mathbf{z}_2 \right)
  \left(\mathbf{s}_2 \cdot  \mathbf{z}_1\right)  \; , \\
 \Phi^{112}(1,2) & = 3\left( \mathbf{s}_1 \cdot \mathbf{t}_{12}(1)\right) (\mathbf{s}_2 \cdot \mathbf{t}_{12}(2)) - <\mathbf{s}_1 , \mathbf{s}_2>  \; , \nonumber \\
                         & = -  \mathbf{s}_1 \cdot \mathbf{s}_2  - \frac{2 +\cos \psi _{12} }{\sin^2 \psi _{12} } \left(\mathbf{s}_1 \cdot  \mathbf{z}_2 \right)
  \left(\mathbf{s}_2 \cdot  \mathbf{z}_1\right)  \; .
 \end{align}
\end{subequations}
In the case of polar fluids only the projections $g^{000}(r_{12})$, $h^{110}(r_{12})$,
and $h^{112}(r_{12})$ have a real physical significance. In    $\mathcal{S}_3(O,R)$ , these functions obviously
depend on the sole $r_{12}=R  \psi_{12}$ at equilibrium, \textit{i.e.} the geodesic distance between the two particles $(1,2)$. 
For a fluid of mono-dipoles $0< \psi _{12}< \pi$, however,  for the fluid of bi-dipoles, only the range
 $0< \psi_{12} < \pi /2 $ is available, because of  the special boundary conditions involved in the model.
The projections $h^{110}(r_{12})$ and  $h^{112}(r_{12})$  are given by~\cite{Caillol_2_a}
\begin{subequations}
\label{h}
  \begin{align}
   h^{110}(r_{12}) &= 3 \int \frac{d\Omega_1}{4 \pi} \; \int \frac{d\Omega_2}{4 \pi} g(1,2)   \Phi^{110}(1,2) \; ,  \nonumber \\
                            &= \frac{3V}{N(N-1)}\Big\langle \sum_{i\neq j=1}^N\frac{\Phi^{110}(1,2) \chi(\psi_{ij}-\psi_{12})}{4\pi R^3 \sin^2(\psi_{ij})\delta \psi} \Big\rangle.\, \\
   h^{112}(r_{12}) &= \frac{3}{2} \int \frac{d\Omega_1}{4 \pi} \; \int \frac{d\Omega_2}{4 \pi} g(1,2)   \Phi^{112}(1,2) \; ,  \nonumber \\
&= \frac{3V}{2N(N-1)}\Big\langle \sum_{i\neq j=1}^N\frac{\Phi^{112}(1,2) \chi(\psi_{ij}-\psi_{12})}{4\pi R^3 \sin^2(\psi_{ij})\delta \psi} \Big\rangle.\, 
  \end{align}
\end{subequations}
where $\Omega_i$ ($i=1,2$) denotes the spherical coordinates of vector $\mathbf{s}_i$ in the local basis
at point $M_i$, $\delta \psi$ is the bin size and $\chi$ is defined as
\begin{equation}
\chi(\psi-\psi_{12}) = \left\{ \begin{array}{ll} 1 & \mbox{if $\psi_{12} < \psi < \psi_{12}+\delta \psi$} \\
0 & \mbox{otherwise.}
\end{array} \right.
\end{equation}

Clearly the pair potentials can be reexpressed in term of the invariants  $\Phi^{110}(1,2)$ and  $\Phi^{112}(1,2) $. One checks that
\begin{subequations}
  \begin{align}
  W_{\boldsymbol{\mu}_1 ,\boldsymbol{\mu}_2}^{\mathrm{mono}} =& \frac{\mu^2}{3 \pi R^3}
                                                                              \left[
                                                                             f\left(\psi_{12}\right) \left(2  \Phi^{112}(1,2)  -   \Phi^{110}(1,2)\right) \right. \nonumber \\
                                                               &  \left.      -2\left(1 + \cos\psi_{12}\right) \left(\Phi^{112}(1,2)  +   \Phi^{110}(1,2)\right) 
 \right]                   \, ,          \\
 W_{\boldsymbol{\mu}_1 ,\boldsymbol{\mu}_2}^{\mathrm{bi}} =&   \frac{\mu^2}{3 \pi R^3}
 \left[ 2\left(1-\cos \psi_{12} \right)  \Phi^{000}(1,2) -\left(1+ 2\cos \psi_{12} \right)  \Phi^{112}(1,2)
\right]  \; .
  \end{align}
\end{subequations}

%%%%%%%%%%%%%%%%%%%%%%%%%%%%%%%%%%%%%%%%%%
%%%%%%%%%%%%%%%%%%%%%%%%%%%%%%%%%%%%%%%%%%
\begin{figure}[t!]
\includegraphics[scale=0.95]{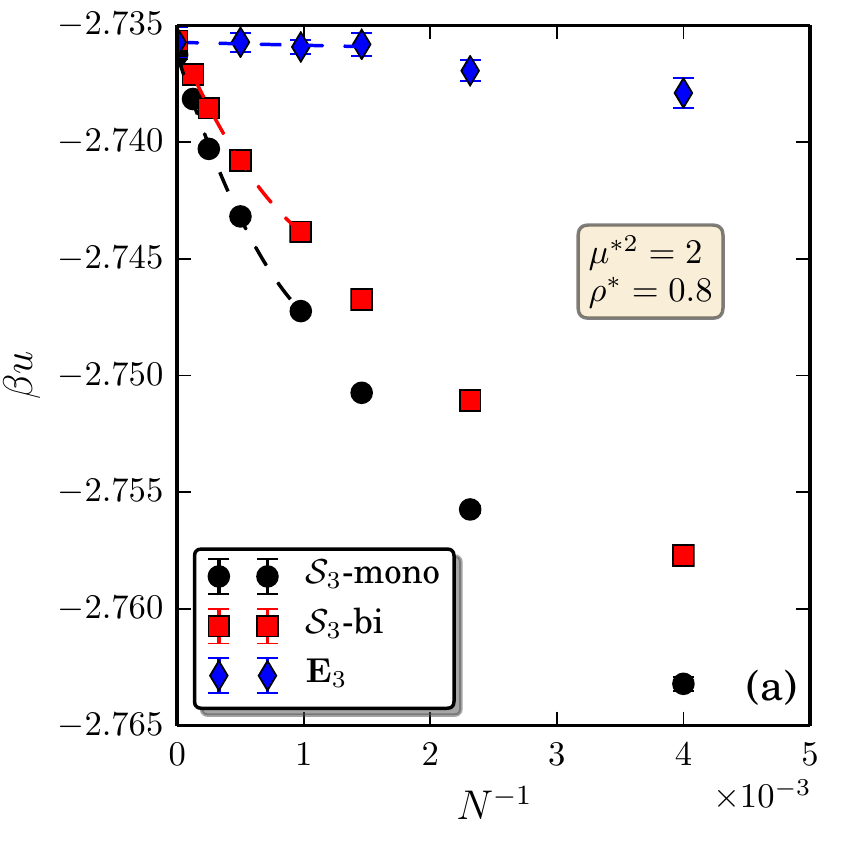}
\includegraphics[scale=0.95]{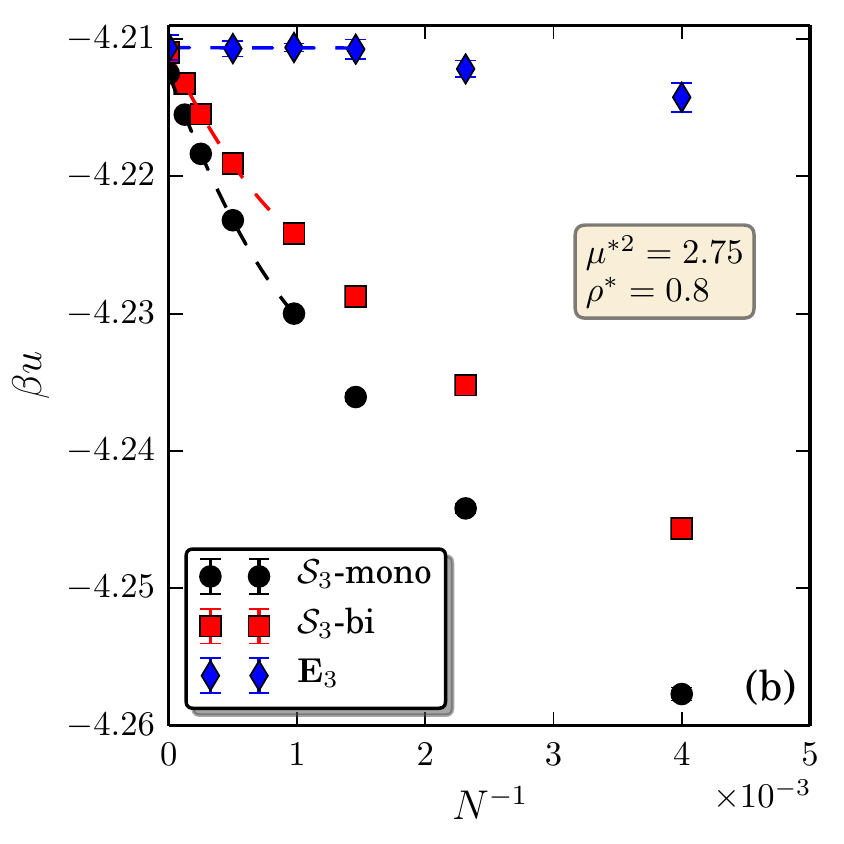}
\includegraphics[scale=0.95]{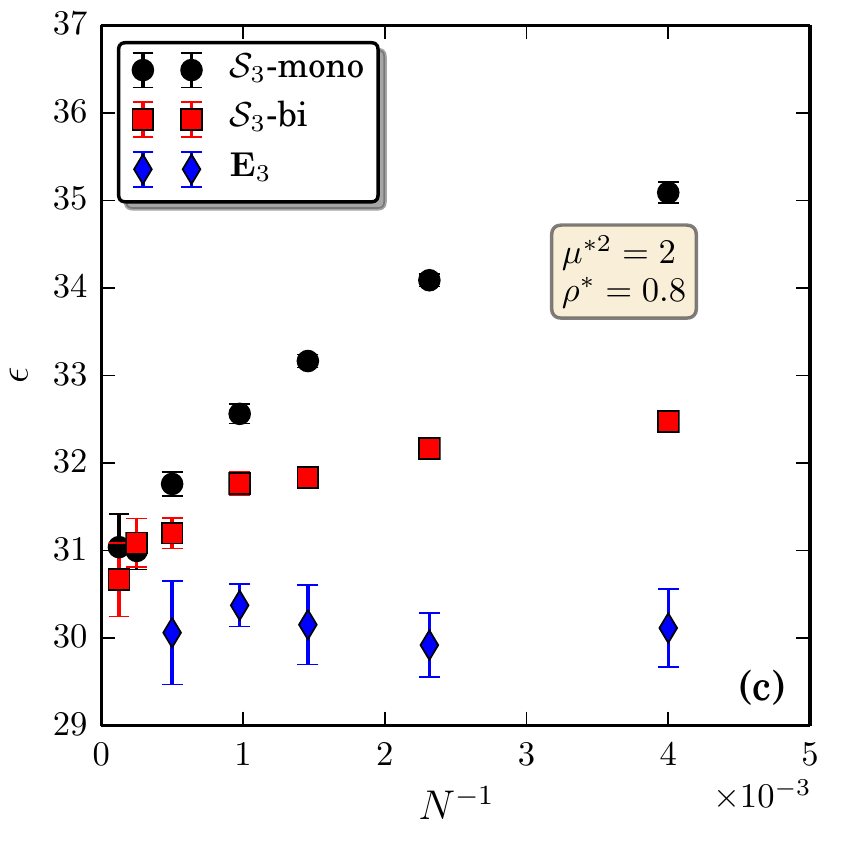}
\includegraphics[scale=0.95]{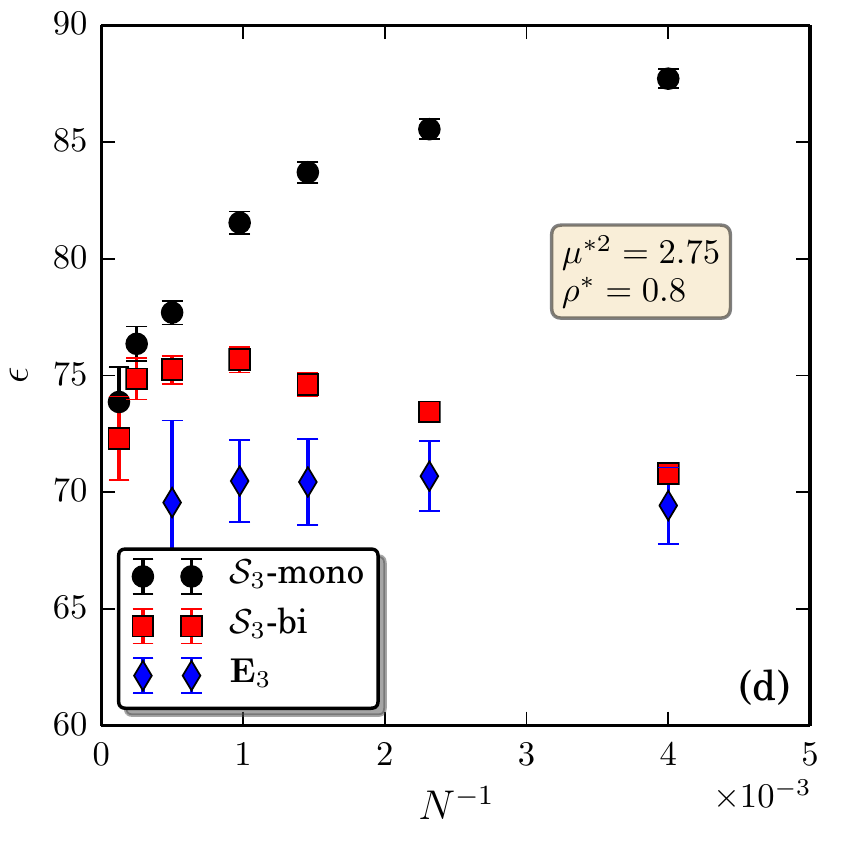}
\caption{Size convergence of both  energies $\beta u$ and  dielectric constant $\epsilon$ for  two
considered thermodynamics  states  and for the three different potentials. Error bars
correspond to two standard deviations. Dashed lines display a second/first order least square fit of the four/three largest systems ($\mathcal{S}_3/\mathbf{E}_3$).}
\label{Fig1}
\end{figure}

%%%%%%%%%%%%%%%%%%%%%%%%%%%%%%%%%%%%%%%%%%
%%%%%%%%%%%%%%%%%%%%%%%%%%%%%%%%%%%%%%%%%%
%%%%%%%%%%%%%%%%%%%%%%%%%%%%%%%%%%%%%%%%%%
%%%%%%%%%%%%%%%%%%%%%%%%%%%%%%%%%%%%%%%%%%
%%%%%%%%%%%%%%%%%%%%%%%%%%%%%%%%%%%%%%%%%%
%%%%%%%%%%%%%%%%%%%%%%%%%%%%%%%%%%%%%%%%%%

\begin{figure}[t!]
\includegraphics[scale=0.95]{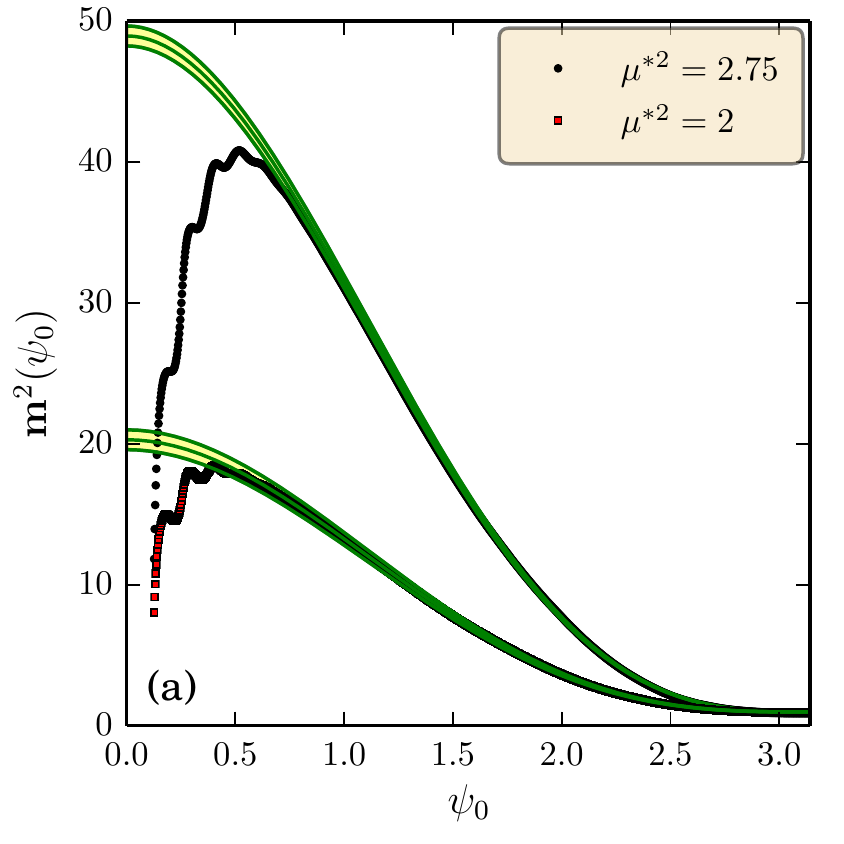}
\includegraphics[scale=0.95]{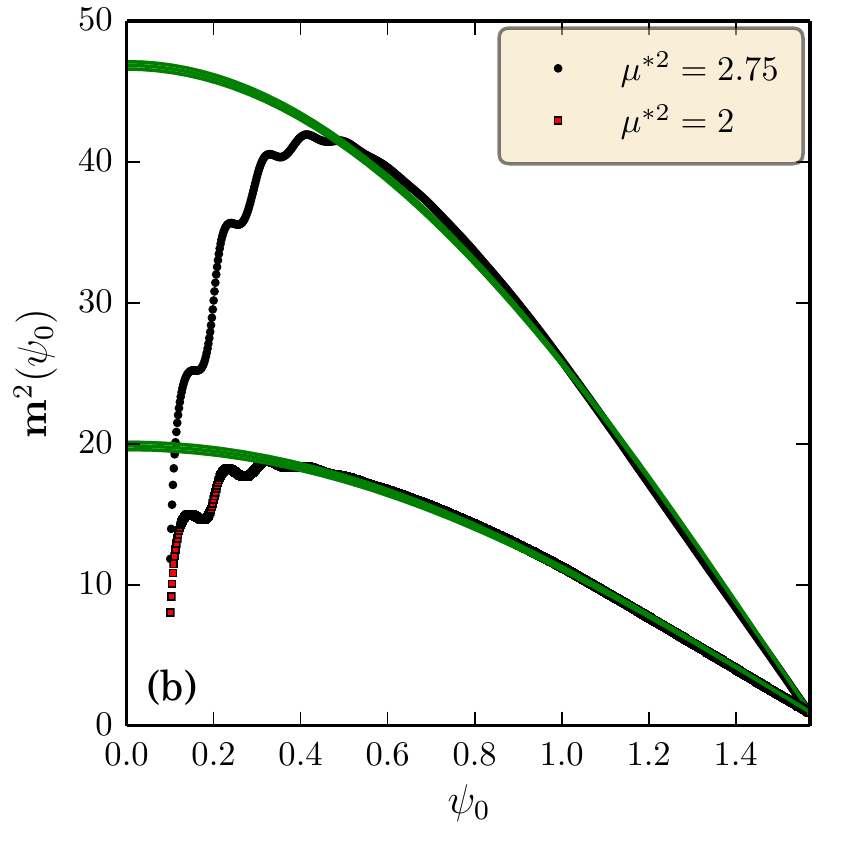}
\caption{Dipolar fluctuations $\boldsymbol{m}^2$ as a function of angle $\psi_0$ for (a) mono-dipoles and (b) bi-dipoles. MC
data are presented with markers while lines correspond to the analytic functions given by Eqs. \eqref{cloc} and \eqref{cloc_mono}. The lines
are shown  for $\epsilon$ values taken from Tables \ref{tab:1/tc} and \ref{tab:3/tc} with their upper and lower bounds.}
\label{Fig2}
\end{figure}

%%%%%%%%%%%%%%%%%%%%%%%%%%%%%%%%%%%%%%%%%%
%%%%%%%%%%%%%%%%%%%%%%%%%%%%%%%%%%%%%%%%%%
%%%%%%%%%%%%%%%%%%%%%%%%%%%%%%%%%%%%%%%%%%

\begin{figure}[t!]
\includegraphics[scale=0.95]{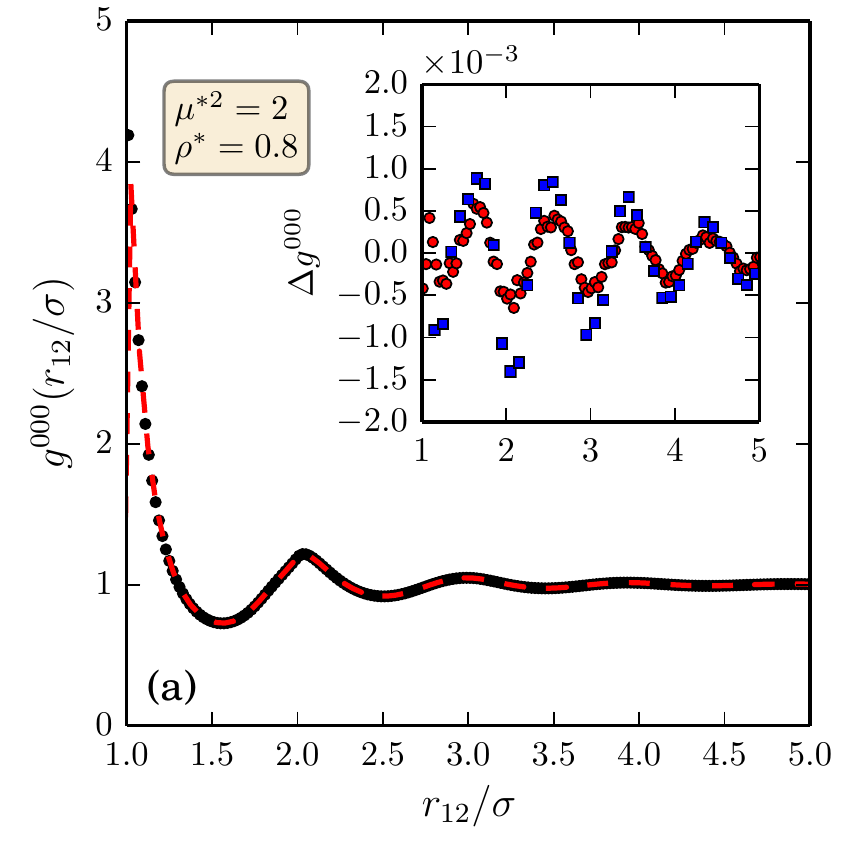}
\includegraphics[scale=0.95]{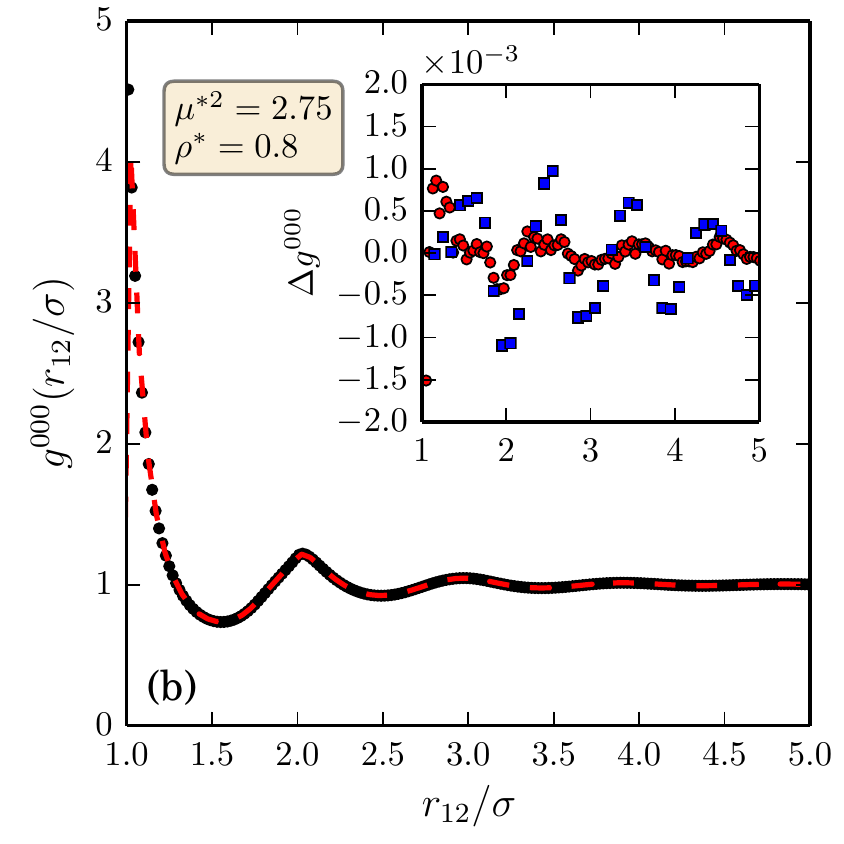}

\caption{Radial pair correlation functions $g^{000}$ for 8000 bi-dipoles $\mathcal{S}_3$-bi (black circles) and Ewald $\mathbf{E}_3$ 
(dashed red line) at $\rho^*=0.8$ and for (a) $\mu^{*2}=2$ and (b)  $\mu^{*2}=2.75$. \textit{Insets:}
show the differences ($\Delta g^{000}$) between $\mathcal{S}_3$-bi and $\mathcal{S}_3$-mono dipoles (red circles) and between $\mathbf{E}_3$
 and $\mathcal{S}_3$-mono dipoles (blue squares).}
\label{Fig3}
\end{figure}

%%%%%%%%%%%%%%%%%%%%%%%%%%%%%%%%%%%%%%%%%%
%%%%%%%%%%%%%%%%%%%%%%%%%%%%%%%%%%%%%%%%%%

\begin{figure}[t!]
\includegraphics[scale=0.95]{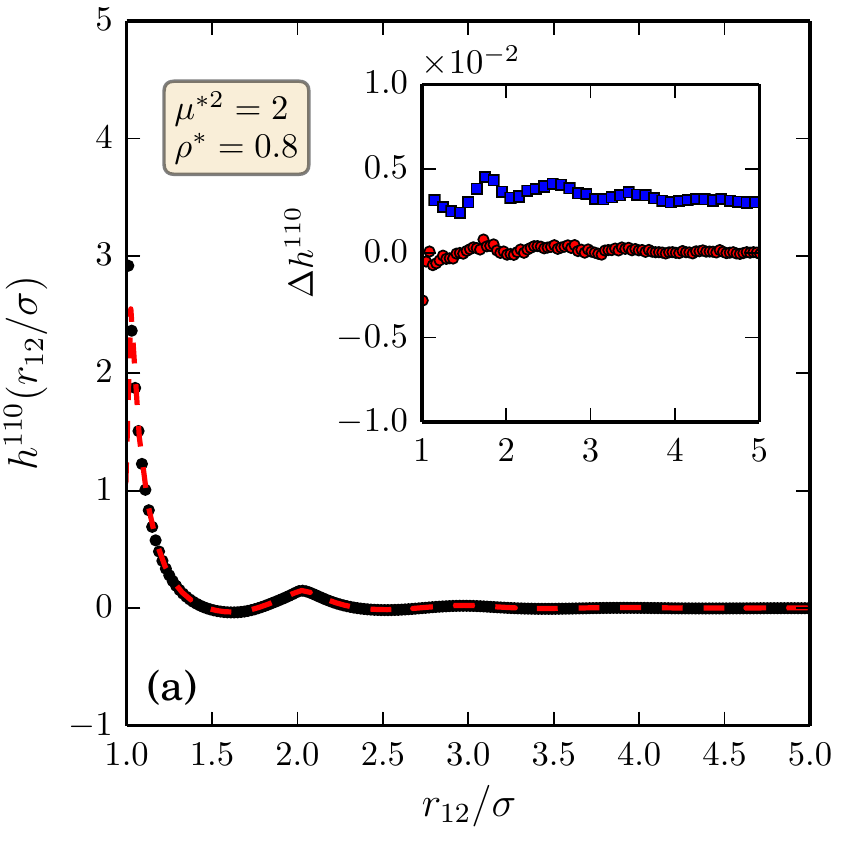}
\includegraphics[scale=0.95]{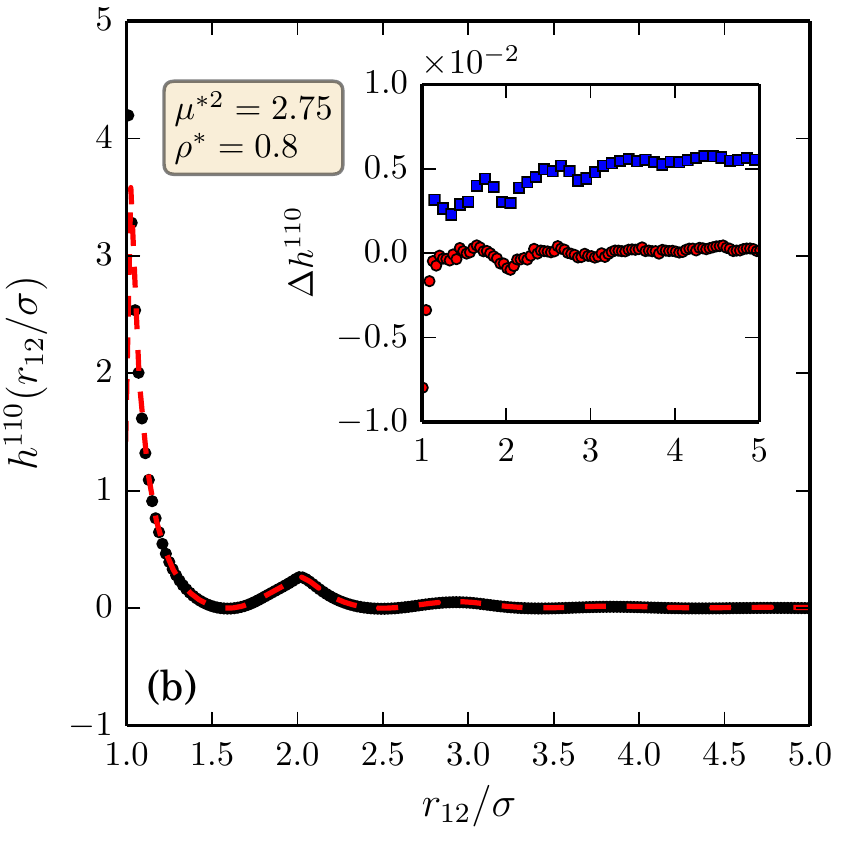}
\includegraphics[scale=0.95]{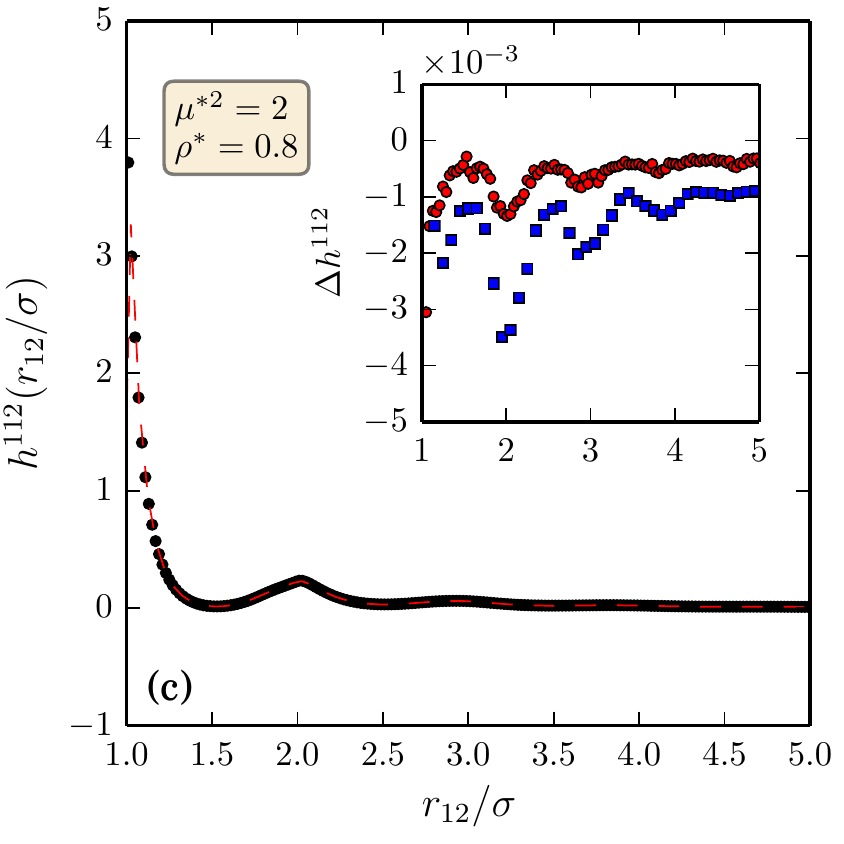}
\includegraphics[scale=0.95]{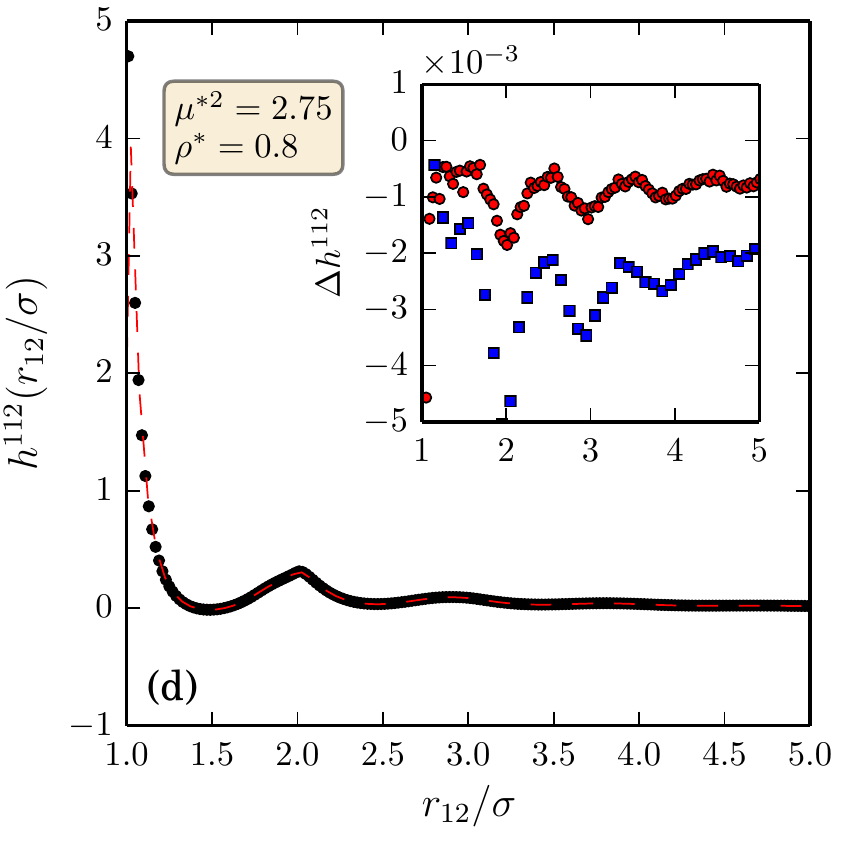}
\caption{Rotational invariants $h^{110}$ and $h^{112}$ for  $\mathcal{S}_3$-bi (black circles) and $\mathbf{E}_3$ (dashed red line) at the system size $N=8000$ for the
two systems. \textit{Insets:} show the differences $\Delta h^{110/112}$ between $\mathcal{S}_3$-bi and $\mathcal{S}_3$-mono and (red circles) and between $\mathbf{E}_3$ and $\mathcal{S}_3$-mono (blue squares).}
\label{Fig4}
\end{figure}
%%%%%%%%%%%%%%%%%%%%%%%%%%%%%%%%%%%%%%%%%%
%%%%%%%%%%%%%%%%%%%%%%%%%%%%%%%%%%%%%%%%%%

\begin{table}[t!]
\centering
\begin{ruledtabular}
\begin{tabular}{||c|c|c|c|c|c||}
\hline
$N$ & $\beta u^{\rm mono}$ & $\beta u^{\rm bi}$ & $\epsilon^{\rm mono}$ & $\epsilon^{\rm bi}$ & \#configs\\
\hline
128 & -2.77588 $\pm$ 4.3e-4 & -2.76979 $\pm$ 4.9e-4 & 36.41 $\pm$ 0.12 & 31.46 $\pm$ 0.10 & 1.3$\times$10$^9$\\
\hline
250 & -2.76321 $\pm$ 3.1e-4 & -2.75771 $\pm$ 2.8e-4 & 35.09 $\pm$ 0.12 & 32.47 $\pm$ 0.11 & 2.5$\times$10$^9$\\
\hline
432  & -2.75574 $\pm$ 1.5e-4 & -2.75106 $\pm$ 1.2e-4 & 34.09 $\pm$ 0.07 & 32.17$\pm$ 0.09 & 12.1$\times$10$^9$\\
\hline
686  & -2.75074 $\pm$ 1.1e-4 & -2.74675 $\pm$ 1.0e-4 & 33.17 $\pm$ 0.07 &  31.84$\pm$ 0.09 & 22.0$\times$10$^9$\\
\hline
1024 & -2.74724 $\pm$ 1.2e-4 & -2.74385 $\pm$ 1.2e-4 & 32.56 $\pm$ 0.11 & 31.77 $\pm$ 0.13 & 16.4$\times$10$^9$\\
\hline
2000 & -2.74319 $\pm$ 1.0e-4 & -2.74078 $\pm$ 1.3e-4 & 31.76 $\pm$ 0.14 & 31.20 $\pm$ 0.18 & 16.0$\times$10$^9$\\
\hline
4000 & -2.74029 $\pm$ 1.1e-4 & -2.73855 $\pm$ 1.1e-4 & 31.00 $\pm$ 0.21 & 1.09 $\pm$ 0.28 & 16.0$\times$10$^9$\\
\hline
8000 & -2.73816 $\pm$ 1.3e-4 &  -2.73712 $\pm$ 1.1e-4 & 31.04 $\pm$ 0.38 &  30.67 $\pm$ 0.42 & 16.0$\times$10$^9$\\
\hline
$\infty$ & -2.73626 $\pm$ 2.0e-4 &  -2.73567 $\pm$ 1.9e-4  & - & - & -\\
\hline
\end{tabular}
\end{ruledtabular}
\caption{\label{tab:1/tc} Number of particles, reduced energies per particle, dielectric constant, and  number of configurations for mono- and bi-dipoles ($\mathcal{S}_3)$ with electrostatic coupling $\mu^{*2}=2$. 
Reported data are given with two standard deviations. TL data are extrapolated via a second order polynomial in $N^{-1}$ from the four largest systems.}
\end{table}

%%%%%%%%%%%%%%%%%%%%%%%%%%%%%%%%%%%%%%%%%%
%%%%%%%%%%%%%%%%%%%%%%%%%%%%%%%%%%%%%%%%%%

\begin{table}[t!]
\centering
\begin{ruledtabular}
\begin{tabular}{||c|c|c|c||}
\hline
$N$ & $\beta u$ & $\epsilon$ & \#configs \\
\hline
128 & -2.74305$\pm$ 7.6e-4 & 29.83 $\pm$ 0.29 & 3.8$\times$10$^8$\\
\hline
250 & -2.73790 $\pm$ 6.3e-4 & 30.12$\pm$ 0.44 & 7.5$\times$10$^8$ \\
\hline
432  & -2.73694 $\pm$ 4.4e-4 & 29.92$\pm$ 0.36 & 1.3$\times$10$^9$ \\
\hline
686 & -2.73582 $\pm$ 4.8e-4 & 30.15$\pm$ 0.44 & 2.1$\times$10$^9$\\
\hline
1024 & -2.73593 $\pm$ 3.0e-4 & 30.37 $\pm$ 0.24 & 3.1$\times$10$^9$\\
\hline
2000 & -2.73573 $\pm$ 4.0e-4 & 30.06 $\pm$ 0.59 & 2.0$\times$10$^9$ \\
\hline
$\infty$ & -2.73573 $\pm$ 6.5e-4 & - & -\\
\hline
\end{tabular}
\end{ruledtabular}
\caption{\label{tab:2/tc} Same as in table \ref{tab:1/tc} for  the Ewald potential ($\mathbf{E}_3$).
TL data is extrapolated via a linear fit in $N^{-1}$ from the three largest systems.}
\end{table}

%%%%%%%%%%%%%%%%%%%%%%%%%%%%%%%%%%%%%%%%%%
%%%%%%%%%%%%%%%%%%%%%%%%%%%%%%%%%%%%%%%%%%

\begin{table}[t!]
\centering
\begin{ruledtabular}
\begin{tabular}{||c|c|c|c|c|c||}
\hline
$N$ & $\beta u^{\rm mono}$ & $\beta u^{\rm bi}$ & $\epsilon^{\rm mono}$ & $\epsilon^{\rm bi}$ & \#configs\\
\hline
128 & -4.27856 $\pm$ 6.7e-4 & -4.26387 $\pm$ 6.3e-4 & 85.39 $\pm$ 0.33 & 62.27 $\pm$ 0.26 & 1.5$\times$10$^9$\\
\hline
250  & -4.25770 $\pm$ 4.7e-4 & -4.24565 $\pm$ 4.6e-4 & 87.71 $\pm$ 0.41 & 70.80 $\pm$ 0.35 & 3.0$\times$10$^9$\\
\hline
432  & -4.24418 $\pm$ 3.5e-4 & -4.23521 $\pm$ 3.5e-4 & 85.55 $\pm$ 0.43 & 73.44 $\pm$ 0.43 & 5.2$\times$10$^9$\\
\hline
686 & -4.23607 $\pm$ 2.8e-4 & -4.22876 $\pm$ 2.7e-4 & 83.71 $\pm$ 0.45 & 74.62 $\pm$ 0.49 & 8.2$\times$10$^9$\\
\hline
1024 & -4.23000 $\pm$ 2.2e-4 & -4.22416 $\pm$ 2.2e-4 & 81.54 $\pm$ 0.48 & 75.68 $\pm$ 0.56 & 12.3$\times$10$^9$\\
\hline
2000 & -4.22320 $\pm$ 1.6e-4 & -4.21908 $\pm$ 1.6e-4 & 77.69 $\pm$ 0.50 & 75.25 $\pm$ 0.59 & 24.0$\times$10$^9$\\
\hline
4000 & -4.21836 $\pm$ 1.6e-4 & -4.21547 $\pm$ 1.6e-4 & 76.36 $\pm$ 0.74 &  74.85 $\pm$ 0.88 & 24.0$\times$10$^9$\\
\hline
8000 & -4.21550 $\pm$ 1.9e-4 &  -4.21323 $\pm$ 2.2e-4 & 73.86 $\pm$ 1.49 & 72.30 $\pm$ 1.80 & 13.9$\times$10$^9$\\
\hline
$\infty$ & -4.21251 $\pm$ 2.9e-4 &  -4.21098 $\pm$ 3.2e-4 & - & - & -\\
\hline
\end{tabular}
\end{ruledtabular}
\caption{\label{tab:3/tc} Same as in Table \ref{tab:1/tc} but for $\mu^{*2}=2.75$.}
\end{table}

%%%%%%%%%%%%%%%%%%%%%%%%%%%%%%%%%%%%%%%%%%
%%%%%%%%%%%%%%%%%%%%%%%%%%%%%%%%%%%%%%%%%%

\begin{table}[t!]
\centering
\begin{ruledtabular}
\begin{tabular}{||c|c|c|c||}
\hline
$N$ & $\beta u$ & $\epsilon$ & \#configs \\
\hline
128 & -4.22193 $\pm$ 16.1e-4 & 67.84 $\pm$ 1.40 & 3.8$\times$10$^8$\\
\hline
250 & -4.21424 $\pm$ 10.0e-4 & 69.43$\pm$ 1.63 & 7.5$\times$10$^8$ \\
\hline
432 & -4.21218 $\pm$ 6.1e-4 & 70.69$\pm$ 1.49 & 1.3$\times$10$^9$ \\
\hline
686 & -4.21074 $\pm$ 7.1e-4 & 70.44$\pm$ 1.84 & 2.1$\times$10$^9$\\
\hline
1024 & -4.21062 $\pm$ 2.9e-4 & 70.48 $\pm$ 1.77 & 3.1$\times$10$^9$\\
\hline
2000 & -4.21070 $\pm$ 5.7e-4 & 69.56 $\pm$ 3.51 & 2.0$\times$10$^9$ \\
\hline
$\infty$ & -4.21063 $\pm$ 9.2e-4 & - & -\\
\hline
\end{tabular}
\end{ruledtabular}
\caption{\label{tab:4/tc} Same as Table \ref{tab:2/tc} but for $\mu^{*2}=2.75$.}
\end{table}

%%%%%%%%%%%%%%%%%%%%%%%%%%%%%%%%%%%%%%%%%%
%%%%%%%%%%%%%%%%%%%%%%%%%%%%%%%%%%%%%%%%%%

%%%%%%%%%%%%%%%%%%%%%%%%%%%%%%%%%%%%%%%%%%
%%%%%%%%%%%%%%%%%%%%%%%%%%%%%%%%%%%%%%%%%%
\subsection{Fulton's theory}
\label{Epsilon}
The theory of the dielectric constant of a polar fluid in  $\mathcal{S}_3(O,R)$ was obtained
in Ref.~\cite{Caillol_4} for the fluid of mono-dipoles. We extend this theory 
to the fluid of bi-dipoles  in this Sec.
As in Ref.~\cite{Caillol_4} we work in the framework of Fulton's theory which
realizes a synthesis between  linear response theory of dielectric media and 
electrodynamics~\cite{Fulton}. We consider a fluid of
 $N$ bi-dipoles in  $\mathcal{S}_3^+(O,R)$ at thermal equilibrium in the presence 
of an external electrostatic field $\boldsymbol{\mathcal{E}}(M)\in \mathcal{T}(M) $. This field is created for instance 
by a static distribution of bi-charges. The medium then acquires a macroscopic
polarization 
\begin{equation}
 \mathbf{P} (M) = < \widehat{\mathbf{P}} (M)>_{\boldsymbol{\mathcal{E}}}  \; ,
\end{equation}
where the brackets denote the equilibrium average of the dynamical variable
$\widehat{\mathbf{P}} (M)$ in the presence of the external field    $\boldsymbol{\mathcal{E}}$. 
The microscopic polarization $\widehat{\mathbf{P}} (M)$ is defined in $\mathcal{S}_3(O,R)^+$  as
\begin{equation}
\widehat{\mathbf{P}} (M) = \sum_{j=1}^N   \mathbf{U}_{\mathcal{S}_3}(\mathbf{z})  \cdot
\boldsymbol{\mu}_j \delta(M,M_j) \; .
\end{equation}
The relation between the macroscopic polarization $\mathbf{P}$ and the external field 
 $\boldsymbol{\mathcal{E}}(M) $ can be established in the framework of
linear-response theory (provided that  $\boldsymbol{\mathcal{E}}(M)$
is small enough) with the result
\begin{equation}
\label{LRT}
 4 \pi \mathbf{P} = \boldsymbol{\chi} \circ  \boldsymbol{\mathcal{E}} \left(\equiv 
 \int_{\mathcal{S}_3(O,R)^{+}}  d  \tau (M^{'})\; \boldsymbol{\chi} (M,M^{'})  \cdot \boldsymbol{\mathcal{E}}(M^{'}) \right) \; .
\end{equation}
The r.h.s. of Eq.~\eqref{LRT} has been formulated in  a compact, albeit convenient notation that will
be adopted henceforth,  where
the symbol $\circ$ means both a tensorial contraction (denoted by the dot '' $\cdot $ '')
and a spacial convolution over the whole space (here $\mathcal{S}_3(O,R)^+$).
From standard linear response theory the susceptibility $ \boldsymbol{\chi} $ is  given by
\begin{equation}
\label{chi}
\boldsymbol{\chi} (M,M^{'}) = 4 \pi \beta <\widehat{\mathbf{P}} (M) \widehat{\mathbf{P}} (M^{'})> \; ,
\end{equation}
where the thermal average $< \ldots >$ in the r.h.s. of~\eqref{chi} are evaluated in the absence of the external 
field and $\beta =1/k_BT$.
The dielectric properties of the fluid are characterized by the dielectric constant $\epsilon$
which however is described in a slightly different way than  $\boldsymbol{\chi}$,
according to the constitutive  relation
\begin{equation}
\label{Max}
  4 \pi \mathbf{P}= (\boldsymbol{\epsilon} - \mathbf{I}) \circ \mathbf{E} \; ,
\end{equation}
where $ \mathbf{E}$ denotes the Maxwell field and $\mathbf{I}(M,M^{'}) \equiv 
\mathbf{U}_{\mathcal{S}_3}(\mathbf{z}) \delta(M,M^{'})$.
In Eq.~\eqref{Max} the Maxwell field $ \mathbf{E}$ is the sum of the external
field  $\boldsymbol{\mathcal{E}}(M)$ and the induced field created by the macroscopic
polarization $ \mathbf{P}$. It is generally assumed that $\boldsymbol{\epsilon}$ is a local function,
\textit{i.e.}  $\boldsymbol{\epsilon}= \epsilon \mathbf{I}$. 
More precisely, it is plausible -and we shall take it for granted- that $\boldsymbol{\epsilon}(M,M^{'})$
is a short range function of the distance between the two points $(M,M^{'})$, at least for a homogeneous liquid,
and one then  defines
\begin{equation}
 \epsilon \mathbf{U}_{\mathcal{S}_3}(\mathbf{z})=
 \int_{\mathcal{S}_3(O,R)^{+}}  d  \tau (M^{'}) \;  \boldsymbol{\epsilon}(M,M^{'})
\end{equation}

Obviously one has
\begin{equation}
\label{zon}
 \mathbf{E} =  \boldsymbol{\mathcal{E}} + 4 \pi \mathbf{G}_0 \circ \mathbf{P} \; ,
\end{equation}
where  $\mathbf{G}_0(M,M^{'})$ is the dipolar Green's function~\eqref{G0}. In general
$(\boldsymbol{\epsilon}- \mathbf{I}) \neq \boldsymbol{\chi} $ since the Maxwell field  $ \mathbf{E}(M)$
and the external field  $\boldsymbol{\mathcal{E}}(M)$
do not coincide. The relation between the two fields is easily obtained from~\eqref{zon}
and usually recast as~\cite{Fulton,Caillol_4}
\begin{equation}
\label{Ee}
  \mathbf{E} =   \boldsymbol{\mathcal{E}} + \mathbf{G} \circ \boldsymbol{\sigma}  \circ  
 \boldsymbol{\mathcal{E}} \; ,
\end{equation}
where $\boldsymbol{\sigma} \equiv \boldsymbol{\epsilon} - \mathbf{I} $  and $ \mathbf{G}(M,M^{'})$
is the macroscopic dielectric Green's function defined by the identity
\begin{equation}
\label{G}
 \mathbf{G} =  \mathbf{G}_0 \circ \left(\mathbf{I}- \boldsymbol{\sigma} \circ
\mathbf{G}_0 \right)^{-1} \; .
\end{equation}
To apprehend the physical significancy of  $\mathbf{G}$ let us consider a point dipole
$\boldsymbol{\mu}_0$ located at point $M_0$ of $\mathcal{S}_3(0,R)$. It creates an external field
$\boldsymbol{\mathcal{E}}(M) = 4 \pi \mathbf{G}_0(M,M_0) \cdot \boldsymbol{\mu}_0$. It follows then
from Eq.~\eqref{Ee} that the Maxwell field is given by
\begin{equation}
  \mathbf{E}(M) = 4 \pi \left(  \mathbf{G}_0 + \mathbf{G} \circ \boldsymbol{\sigma} \circ  
 \mathbf{G}_0 \right)(M,M_0) \cdot \boldsymbol{\mu}_0 \;.
\end{equation}
However $ \mathbf{G} \circ \boldsymbol{\sigma} \circ  
 \mathbf{G}_0 = \mathbf{G}_0 \circ 
[\mathbf{I} -\boldsymbol{\sigma}\circ  \mathbf{G}_0  ]^{-1}\circ
 [ \boldsymbol{\sigma} \circ \mathbf{G}_0  -\mathbf{I} + \mathbf{I} ] = - \mathbf{G}_0 +  \mathbf{G} $ from which it follows 
that $ \mathbf{E}(M) = 4 \pi  \mathbf{G}(M,M_0) \cdot \boldsymbol{\mu}_0$ represents
the electric field due to the dipole in the presence of the dielectric medium. Assuming the locality of the dielectric
constant leads us to guess that for $\mathcal{S}_3$, $ \mathbf{G}(M,M_0)= \mathbf{G}_0(M,M_0)/ \epsilon$ (in the absence
of walls).

Combining Eqs.~\eqref{LRT}, ~\eqref{Max}, and ~\eqref{Ee} yields Fulton's relation
\begin{equation}
 \label{F_rel}
\boldsymbol{\chi} = \boldsymbol{\sigma}  +  \boldsymbol{\sigma} \circ  \mathbf{G} \circ \boldsymbol{\sigma} \; .
\end{equation}

To go further one has to compute seriously the macroscopic Green's function $ \mathbf{G}$ and check our guess.
Our starting point
is the following identity, proved in the appendix :
\begin{equation}
 \mathbf{G}_0 \circ  \mathbf{G}_0  = - \mathbf{G}_0  \; .
\end{equation}
Therefore $ - \mathbf{G}_0$ is a projector and has no inverse. Assuming the locality of $\boldsymbol{\sigma}$
one is then led to search the inverse  $\left(\mathbf{I}- \boldsymbol{\sigma} \circ
\mathbf{G}_0 \right)^{-1}$ in the r.h.s. of~\eqref{G} under the form $a \mathbf{I} + b \mathbf{G}_0 $
where $a$ and $b$ are numbers (or local operators). By identification one finds $a=1$ and $b=\sigma/(1+ \sigma)$
yielding for  $ \mathbf{G}$ the simple (and expected) expression
\begin{equation}
  \mathbf{G}= \mathbf{G}_0 /(1+ \sigma) \equiv \mathbf{G}_0/\epsilon \; .
\end{equation}
This results allows to recast Fulton's relation~\eqref{F_rel} under its final form
\begin{equation}
 \label{F_rel_bis} 
(\epsilon-1)  \mathbf{I}(M_1,M_2) + \frac{(\epsilon-1)^2}{\epsilon} \;  \mathbf{G}_0 (M_1,M_2) = \boldsymbol{\chi} (M_1,M_2)  \; .
\end{equation}
We stress that the above equation has been obtained under the assumption of the locality
of the dielectric tensor  $\boldsymbol{\epsilon}(M,M^{'})$. Therefore it should be valid only asymptotically,
\textit{i.e.} for points $(M,M^{'})$ at a mutual distance larger then the range $\xi$ of $\boldsymbol{\epsilon}(M,M^{'})$.
%%%%%%%%%%%%%%%%%%%%%%%%%%%%%%%%%%%%%%%%%%
%%%%%%%%%%%%%%%%%%%%%%%%%%%%%%%%%%%%%%%%%%
\subsection{The dielectric constant and the Kirkwood's factor}
\label{Diel}
Expressions for the dielectric constant, well suited for numerical simulations, can be obtained
from Eq.~\eqref{F_rel_bis} by integration. Slavishly following Refs.~\cite{Berend,Caillol_4} one integrates
both sides of Eq.~\eqref{F_rel_bis} and then takes the trace. The integration of $M_2$ is performed
over a cone of axis $\mathbf{z}_1$ and aperture $\psi_0$ and then $M_1$ is
integrated over  the whole northern hemisphere $\mathcal{S}_3(O,R)^{+}$. The singularity
of the dipolar Green's function  $\mathbf{G}_0(M_1,M_2)$ for $\psi_{12} \sim 0$ must be carefully taken into
account and this delicate point is detailed in the appendix (see Eq.~\eqref{int1}). One finds finally
  \begin{equation}
\label{cloc}
   \frac{\epsilon -1}{\epsilon} + \frac{2}{3} \frac{(\epsilon -1)^2}{\epsilon} \cos \psi_0 = \mathbf{m}^2(\psi_0) \; ,
  \end{equation}
where the dipolar fluctuation $ \mathbf{m}^2(\psi_0)$ reads as
  \begin{equation}
   \mathbf{m}^2(\psi_0) = \frac{4 \pi \beta \mu^2}{3 V}  < \sum_i^N \sum_j^N \mathbf{s}_i \cdot  \mathbf{s}_j \, \Theta (\psi_0 -\psi_{ij})> \; ,
  \end{equation}
where $\Theta(x)$ is the Heaviside step-function ($\Theta(x)=0$ for $x<0$ and  $\Theta(x)=1$ for $x>0$).

We  have thus obtained a family of formula depending on parameter $\psi_0$; clearly they should be valid  only if $R\psi_0$ is large
when compared to the range of the dielectric constant. The numerical results of Sec.~\eqref{Simulations}
show that this range is of the order of a few atomic diameters.
It is also important to note that for $\psi_0=\pi/2$ Eq.~\eqref{cloc} involves the fluctuations
of the total $4 \mathrm{D}$ dipole moment of the system.  However, the resulting formula \textit{i.e.}
$(\epsilon -1)/\epsilon=\mathbf{m}^2(\pi/2) $, albeit simple, 
is not adapted for numerical applications since, for  large values  of the dielectric constant,
a reasonable numerical error on $\epsilon$ requires a determination of
$\mathbf{m}^2(\pi/2) $ with an impractical precision. 
The choice $\psi_0=\pi/3$ yields the less simple formula
$(\epsilon-1) (\epsilon+2)/(3 \epsilon) = \mathbf{m}^2(\pi/3) $ which however allows, by contrast, a precise
determination of $\epsilon$.  Indeed, let $\delta \epsilon$ be the error on $\epsilon$, then, for  high values of the dielectric constant 
the errors on $\mathbf{m}^2(\pi/3)$ and  $ \epsilon$ are roughly  linearly proportional
as $\delta \epsilon\sim3 \,  \delta \mathbf{m}^2(\pi/3)$ 

Note that formula~\eqref{cloc} relating $\epsilon$ to the fluctuation  $ \mathbf{m}^2(\psi_0)$ are similar
but not identical to that obtained for mono-dipoles~\cite{Caillol_4} that we recall below for the sake of completeness :
  \begin{equation}
\label{cloc_mono}
\mathbf{m}^2(\psi_0) =\frac{\epsilon -1}{\epsilon} + \frac{(\epsilon -1)^2}{\epsilon} a(\psi_0) \; ,
 \end{equation}
with 
 \begin{equation}
  a(\psi) = \frac{2}{3 \pi} \left( \sin \psi +(\pi - \psi) \cos \psi \right) \; .
 \end{equation}

The fluctuation $\mathbf{m}^2(\psi_0) $ is of course related to  the Kirkwood factor $g^K(\psi_0)$.
One has  $\mathbf{m}^2(\psi_0) = 3 y g^K(\psi_0)$, with $y=4 \pi \beta \rho \mu^2/9$ and
 \begin{equation}
\label{Kirk}
  g^K(\psi_0) = 1 + \frac{\rho}{3} \, R^3 \int_0^{\psi_0}\; 4 \pi \sin^2 \psi \,  h^{\Delta}(\psi) d \psi \; ,
 \end{equation}
where 
 \begin{equation}
   h^{\Delta}(r=R \psi) =\frac{1}{3}\, (\cos \psi +2) h^{110}(r) + \frac{2}{3}\, (\cos \psi -1) h^{112}(r) \; .
 \end{equation}
It follows from~\eqref{fi} and é\eqref{h} that 
 \begin{equation}
  \label{hdelta}
   h^{\Delta}(r ) = 
                           \frac{3V}{N(N-1)}\Big\langle \sum_{i\neq j=1}^N\frac{(\mathbf{s}_i \cdot \mathbf{s}_j) \;
 \chi(\psi_{ij}-\psi)}{4\pi R^3 \sin^2(\psi_{ij})\delta \psi} \Big\rangle \, .
 \end{equation}
Note that in the thermodynamic limit (TL), \textit{i.e.} fixed $r$ and $R \to \infty$,  one recovers the usual Euclidian expression
of Kirkwood function~\cite{Hansen}
 \begin{equation}
   \label{hdeltainf}
   h^{\Delta}_{\infty}(r ) = 
                           \frac{3V}{N(N-1)}\Big\langle \sum_{i\neq j=1}^N\frac{(\mathbf{s}_i \cdot \mathbf{s}_j) \;
 \chi(r_{ij}-r)}{4\pi r_{ij}^2 \, \delta r} \Big\rangle \, ,
 \end{equation}
where $\delta r =R \delta \psi $.
%%%%%%%%%%%%%%%%%%%%%%%%%%%%%%%%%%%%%%%%%%
%%%%%%%%%%%%%%%%%%%%%%%%%%%%%%%%%%%%%%%%%%
\subsection{Asymptotic behavior of the pair correlation function.}
\label{Asympto}

%%%%%%%%%%%%%%%%%%%%%%%%%%%%%%%%%%%%%%%%%%
%%%%%%%%%%%%%%%%%%%%%%%%%%%%%%%%%%%%%%%%%%

Fulton's relation~\eqref{F_rel_bis} has been used in Ref.~\cite{Caillol_4} to obtain the asymptotic
behavior of the projections $h^{110}(r)$ and  $h^{112}(r)$ of the pair-correlation function $g(1,2)$
of a fluid of mono-dipoles.  The extension of this analysis to a
fluid of bi-dipoles is trivial and will not be detailed here. Following step by step the derivations of
Ref.~\cite{Caillol_4} one easily obtains that, for large $r=R \psi$ and $\psi < \pi/2$,
one should have asymptotically 
\begin{subequations}
\label{assym}
  \begin{align}
   h^{110}_{\text{asymp.}}(r) &  \sim - \frac{(\epsilon-1)^2}{ y \rho \epsilon} \frac{1}{4 \pi R^3\sin^3 \psi }\frac{2(1 -  \cos \psi)}{3}   \; , \\
   h^{112}_{\text{asymp.}}(r) & \sim \frac{(\epsilon-1)^2}{y \rho \epsilon} \frac{1}{4 \pi R^3 \sin^3 \psi  }\frac{1 + 2 \cos \psi}{3} \; .
  \end{align}
 \end{subequations}
We stress that these asymptotic behaviors are valid, even for a finite radius $R$, as soon as $r >>\xi$, where 
$\xi $ denotes the range of the two point dielectric function $\boldsymbol{\epsilon}(1,2)$. Indeed they are easily obtained from
Fulton's relation~\eqref{F_rel_bis} which assumes the locality of $\boldsymbol{\epsilon}(1,2)$. This point is further discussed
and confirmed by the MC simulations presented in
Sec.~\eqref{Simulations}.
It must be stressed that, in the TL limit $R \to \infty$ \textit{and} with $r \gg \xi$ fixed but large, one recovers the expected
Euclidian behavior $ h^{112}_{\text{asymp.}}(r) \sim (\epsilon-1)^2/(4 \pi y \rho \epsilon) \times 1/r^3$ valid for
an infinite system without boundaries at infinity~\cite{Nien,Hoye,Stell,Caillol_4}.
By contrast, in the same limit, one obtains  that $ h^{110}_{\text{asymp.}}(r)\sim  (\epsilon-1)^2/(4 \pi y \rho \epsilon) \times 1/r \times 1/R^2$
which tends to zero for the infinite system for which $R \to \infty$. This behavior is in agreement with the expected short range behavior of 
the projection $ h^{110}(r)$ in the $3D$ infinite Euclidian space~\cite{Nien,Hoye,Stell}.

Let us now discuss the behavior of $h^{\Delta}(r)$. It follows from~\eqref{assym} that for  $r \gg \xi$ one has
\begin{equation}
 \label{asymDel}
 h^{\Delta}_{\text{asymp.}}(r)   \sim -\frac{2}{3}  \frac{(\epsilon-1)^2}{ y \rho \epsilon}  \frac{1}{4 \pi R^3 \sin \psi  } \; .
\end{equation}
As for $ h^{110}(r)$, in the  TL limit,  $h^{\Delta}_{\text{asymp.}}(r) \to 0$ as $R^{-2}$ at given $r$ and 
 $R \to \infty$.

Although the asymptotic tail of  $h^{\Delta}(r )$ tends to zero uniformly in the limit $R \to \infty$,
 its integral over the volume of the cone
of aperture $\psi_0$ in the r.h.s. of Eq.~\eqref{cloc} gives a finite contribution to the Kirkwood function.
Clearly, for large $R$, Eq.~\eqref{Kirk} can be written as
\begin{equation}
\label{zo}
   g^K(\psi_0) =  g^K_{\infty} + \frac{\rho R^3}{3} \int_0^{\psi_0 } h^{\Delta}_{\text{asymp.}}(r)  4 \pi \sin^2(\psi) d\psi \; ,
\end{equation}
where $g^K_{\infty}$ is the Euclidian Kirkwood factor
\begin{equation}
 g^K_{\infty} =\int_0^{\infty} 4 \pi r^2 dr \,  h^{\Delta}_{\infty}(r )  \; ,
\end{equation}
where $h^{\Delta}_{\infty}(r ) $ is the infinite-volume limit of Kirkwood's pair correlations as defined in Eq.~\eqref{hdeltainf}.
Now, inserting the asymptotic behavior~\eqref{asymDel} of $h^{\Delta}(r)$ in Eq.~\eqref{zo} one obtains
\begin{equation}
\label{blo}
 \frac{(\epsilon -1) (2 \epsilon +1)}{\epsilon} = 9 y g^K_{\infty} \; ,
\end{equation}
which is the well-known Kirkwood formula for the dielectric constant of an infinite Euclidian
polar fluid without boundaries at infinity \cite{Hansen}. The above  mechanism to  get rid of the electrostatic
tail of  $h^{\Delta}_{\text{asymp.}}(r )$ in order to obtain the more intrinsic expression~\eqref{blo} of
the dielectric constant, also works for mono-dipoles in $\mathcal{S}_3(0,R)$ or cubico-periodical
geometries for which a similar explicit calculation can  be performed 
(but will not be reported  here due to lack of space). It is likely to  be a general
mechanism for any arbitrary, Euclidian or not, geometries.

%%%%%%%%%%%%%%%%%%%%%%%%%%%%%%%%%%%%%%%%%%
%%%%%%%%%%%%%%%%%%%%%%%%%%%%%%%%%%%%%%%%%%
\section{Monte Carlo Simulations}
\label{Simulations}
We performed standard Metropolis MC simulations of a DHS fluid with single particle
displacements (translation and rotation), where each new configuration is generated by a trial displacement of \emph{one} dipole.
Two different
systems were studied, both with the same reduced particle density $\rho^*=0.8$, but with different reduced dipolar
 couplings,  $\mu^{*2}=2$ and $\mu^{*2}=2.75$. These systems have previously been studied in the literature \cite{Adams,Patey} and 
serve as a good benchmark for any new potential. Both  systems are known to be in the dielectric fluid phase
 \cite{JJ, Holm} (in contrast to a ferroelectric phase). The system
sizes were systematically varied and the energies and dielectric constants  extrapolated to their thermodynamic limits.
Simulations were either performed on the hypersphere $\mathcal{S}_3$ or in the Euclidian space $\mathbf{E}_3$ 
 with cubic periodic boundary
conditions. Interaction potentials for the the mono- and bi-dipoles on $\mathcal{S}_3$ are given by
 Eq. \eqref{dip_mono} and \eqref{dip_bi} while
in $\mathbf{E}_3$ the  dipolar  Ewald summation techniques \cite{JJ,Holm} were applied. The parameters for the dipolar Ewald potential
 were adapted from an automatic scheme for
charged particles \cite{Linse} using a real-space cut-off equal to half the box-length. Systematic tests were performed to ensure that the
resulting energies and dielectric constants were not influenced by the chosen precision of the dipolar Ewald method
(within error bars).
The data presented for the Ewald method were obtained with tinfoil boundary conditions, split parameters $\alpha$ in the range  $[1.150994\sigma^{-1},0.480594\sigma^{-1}]$,
 and with a number of wave-functions in the range $[871,1059]$  for systems between $N=128 \text{ and }2000$ dipoles.

Below we give results from extensive simulations of DHS in $\mathcal{S}_3$ and $\mathrm{E}_3$ geometries
at different system sizes.
MC data for the energy and the dielectric constant are given in Tabs.~I-IV for the two thermodynamic states $(\rho^*=0.8, \mu^{*2}=2)$
and $(\rho^*=0.8, \mu^{*2}=2.75)$ for the three potentials and various number of particles $N$ as well as the extrapolation to $N \to \infty$.

The energies and the dielectric constants  all converge, as expected, to the same values in the thermodynamic limit for all 
 three potentials (see Fig.~(\ref{Fig1}) and Table \ref{tab:1/tc}-\ref{tab:4/tc}).
The energies can be well fitted with $\beta u=\beta u_{\rm \infty} +\mathcal{O}(1/N)$ for the largest system sizes (for our purpose
we used a second order polynomial in $1/N$). We were incapable to perform a similar analysis for the dielectric constant due
to larger error bars but it seems reasonable to assume that the thermodynamic limit is close to the value  obtained  for $N=8000$ particles.
 We found 
 \{$\beta u_{\rm \infty}= -2.736\pm 0.001$, $\epsilon \simeq 30\pm 2$ \} and  \{$\beta u_{\rm \infty}= -4.212\pm
 0.002$, $\epsilon \simeq 70\pm 5$ \} for the two considered states.
These values are considerably more precise than previous studies of the same systems \cite{Adams,Patey} and serve as
an update of these thermodynamic values. From Fig.~(\ref{Fig1}) one finds that the $\mathbf{E}_3$ and
the Ewald summation techniques tends to give faster size convergence (to the TL), which seems 
to have converged both in energy and dielectric constant already at a system sizes around $N\sim700$.

Note that the dielectric
constant in Figure~(\ref{Fig2}) and Tables \ref{tab:1/tc}-\ref{tab:4/tc} are
calculated from Eqs.~\eqref{cloc_mono} and \eqref{cloc} at the specific angles $\psi_0=\pi/2$ for $\mathcal{S}_3$-mono 
and $\psi_0=\pi/3$ for $\mathcal{S}_3$-bi.  However, when  the fluctuations of the dipole moment 
are integrated over  volumes corresponding to other values of the angle $\psi_0$
the same dielectric constant is obtained as soon as  $\psi_0$ is large enough as can be seen   in Fig.~(\ref{Fig2}).
Only at small $\psi_0$, \text{i.e.} for  $R \psi_0 < \xi$ ($\xi$ range of dielectric constant)
do the fluctuations differ from the predictions of macroscopic (local) electrostatics given by Eqs. \eqref{cloc_mono} and \eqref{cloc}, due to
short-ranged molecular structuring and orientally ordering.

 Figs.~(\ref{Fig3}) and~(\ref{Fig4}) display the isotropic correlation functions $g^{000}(r)$ and the projections  $h^{110}(r)$
and $h^{112}(r)$ and show a very good agreement  with very small structural and orientally differences (less than
$10^{-2}$ units) between the two different geometries ($\mathcal{S}_3$ and $\mathbf{E}_3$) at short separations
 for all the three different potentials ($\mathcal{S}_3$-mono, $\mathcal{S}_3$-bi, and $\mathbf{E}_3$) considered here. The slightly larger 
discrepancy between the Ewald potential
compared to the two hypersphere potentials is most likely due the smaller size in the former ($N=2000$ compared to $N=8000$).

\begin{figure}[t!]
\includegraphics[scale=0.95]{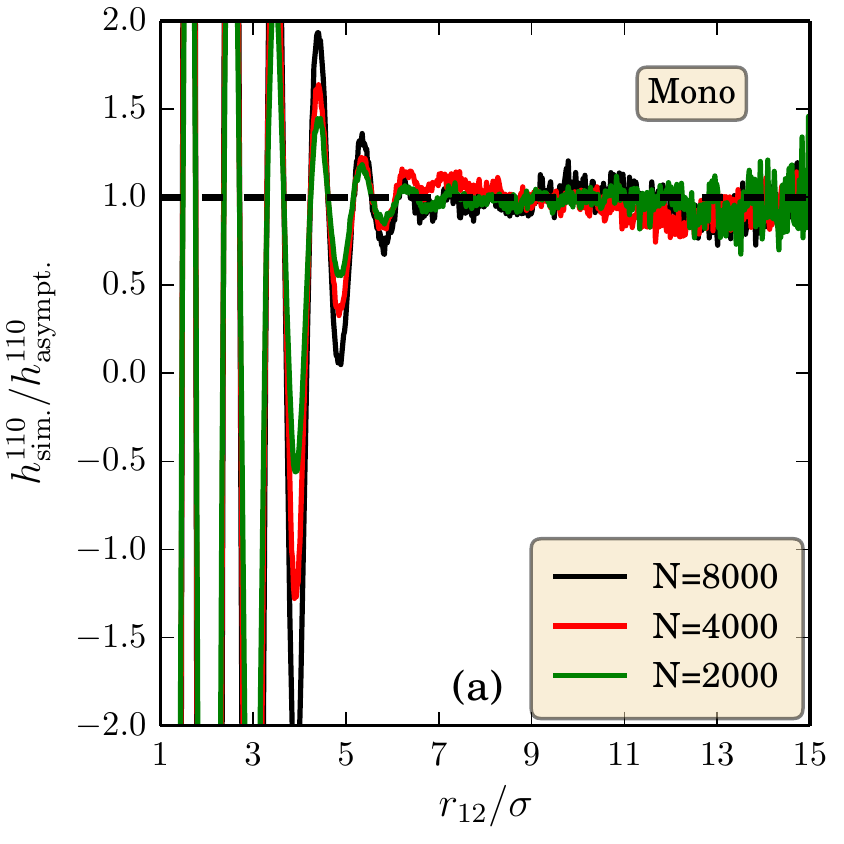}
\includegraphics[scale=0.95]{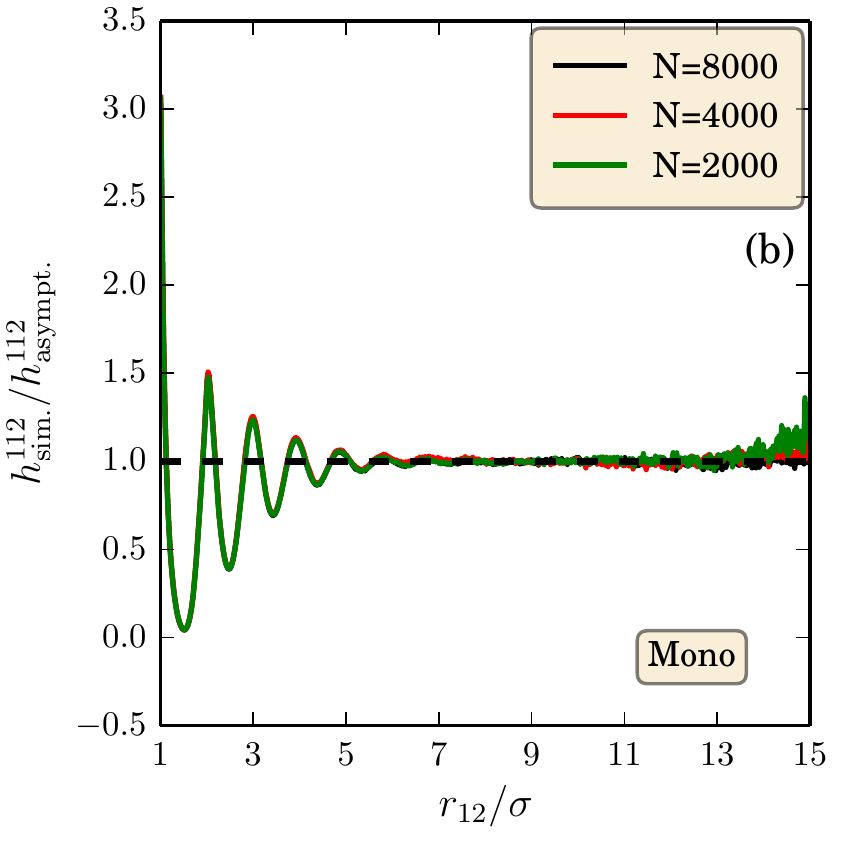}
\includegraphics[scale=0.95]{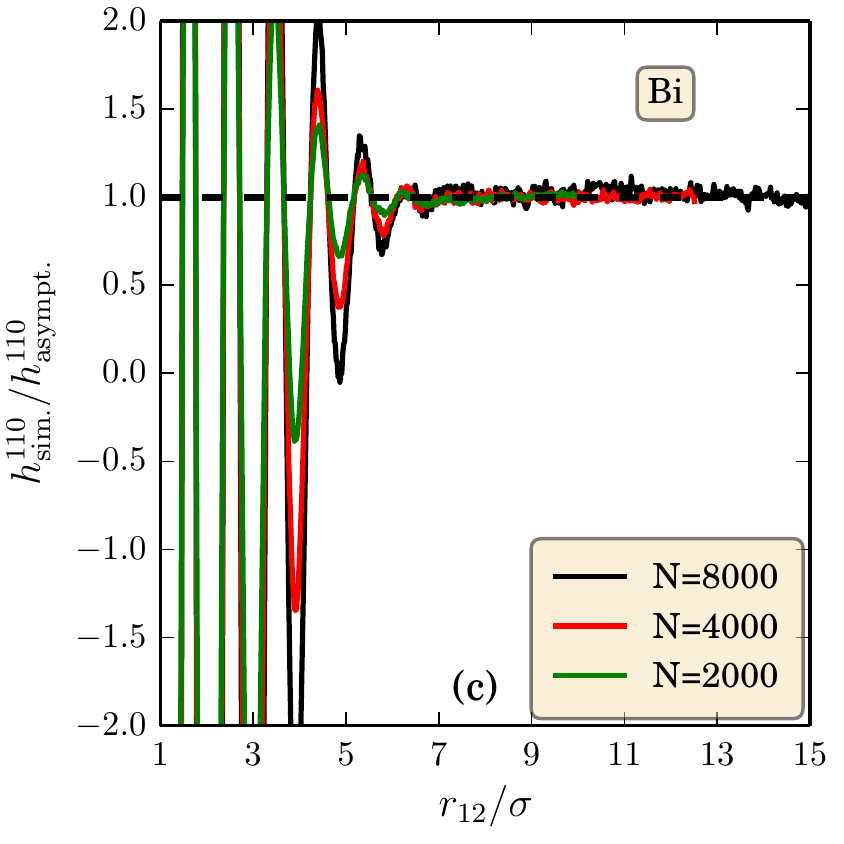}
\includegraphics[scale=0.95]{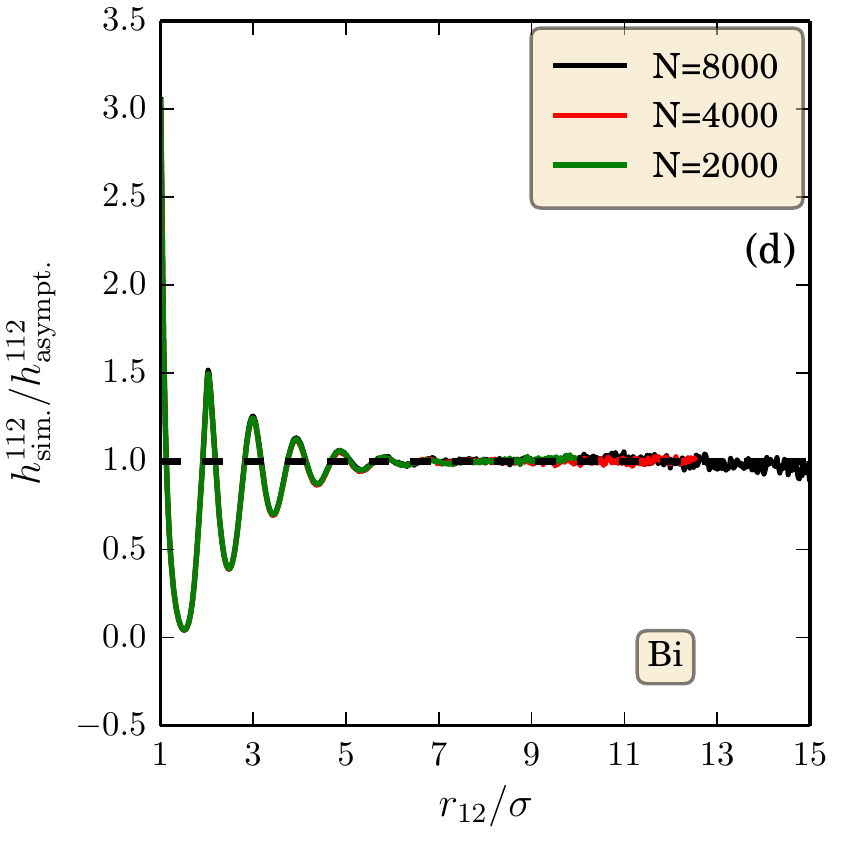}
\caption{Asymptotic behavior of the rotational invariants. 
$h^{110}(r)/h^{110}_{\rm asympt.}(r)$ and
 $h^{112}(r)/h^{112}_{\rm asympt.}(r)$ for (a,b) $\mathcal{S}_3$-mono and (c,d) $\mathcal{S}_3$-bi for 
$\mu^{*2}=2$. Lines
as in Fig. \ref{Fig4} but for the Eqs. \eqref{assym} in this paper and Eqs. (4.32) from \cite{Caillol_4}.}
\label{Fig5}
\end{figure}

\begin{figure}[t!]
\includegraphics[scale=0.95]{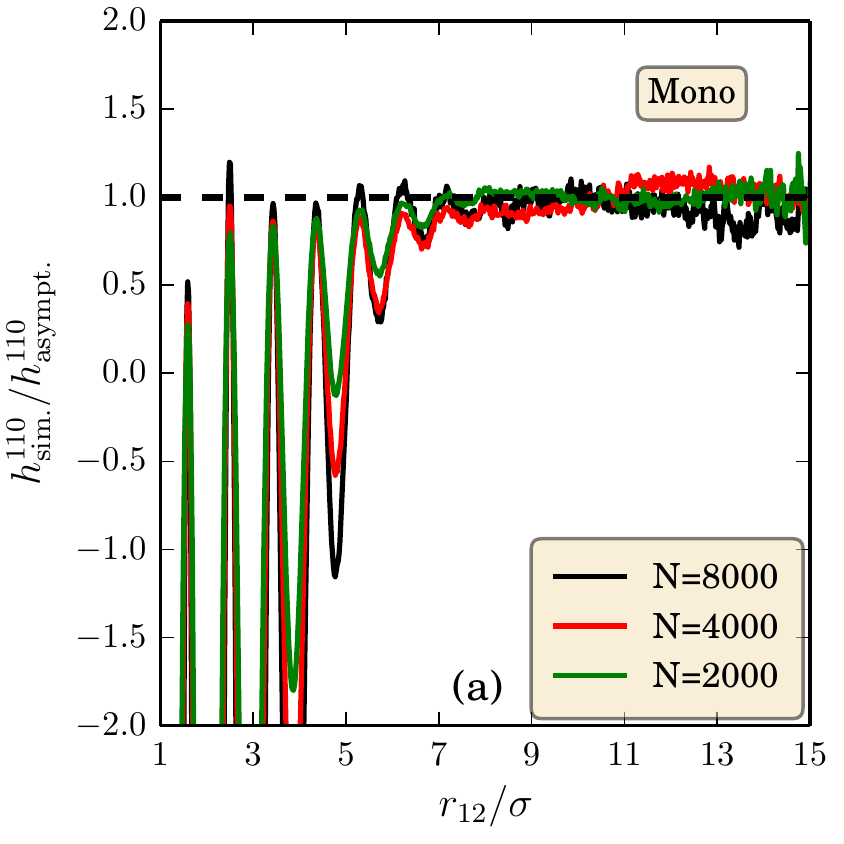}
\includegraphics[scale=0.95]{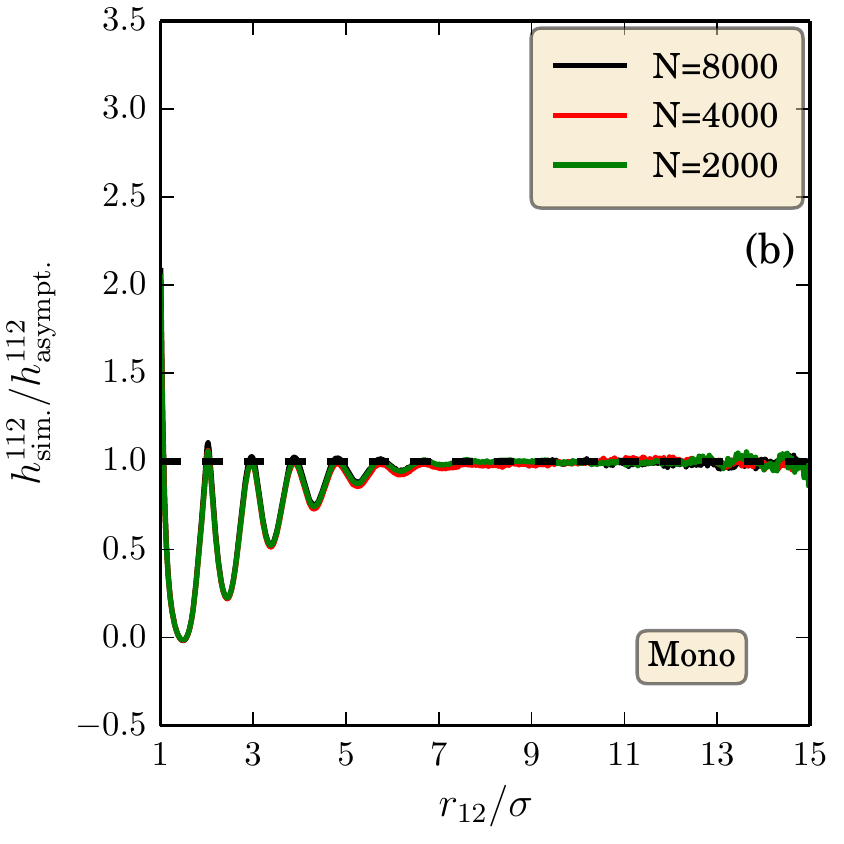}
\includegraphics[scale=0.95]{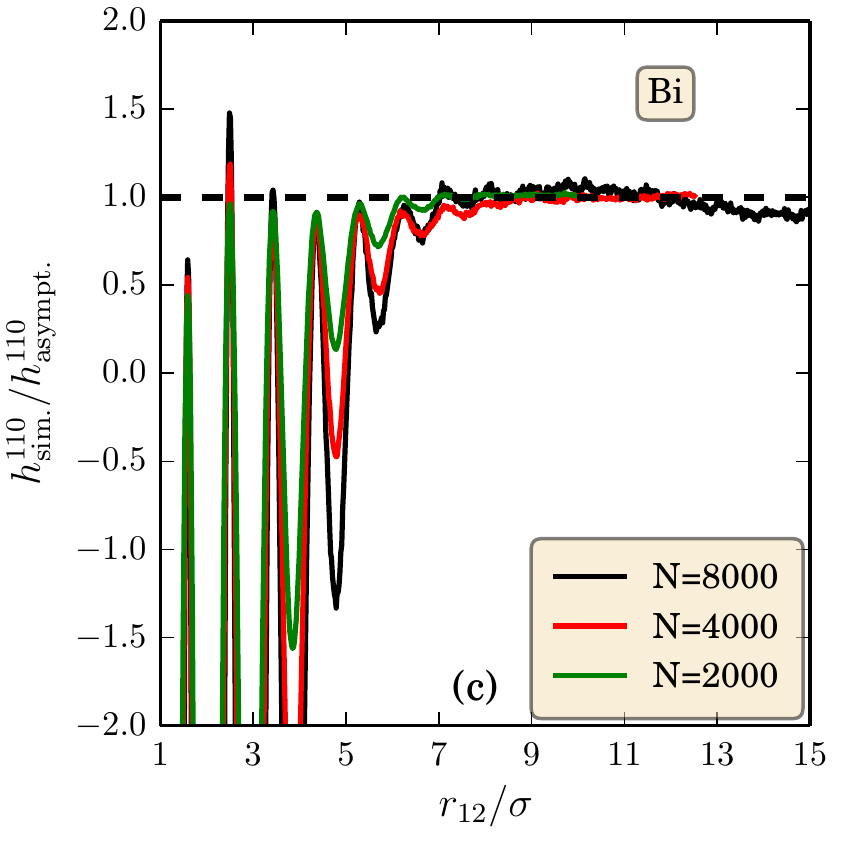}
\includegraphics[scale=0.95]{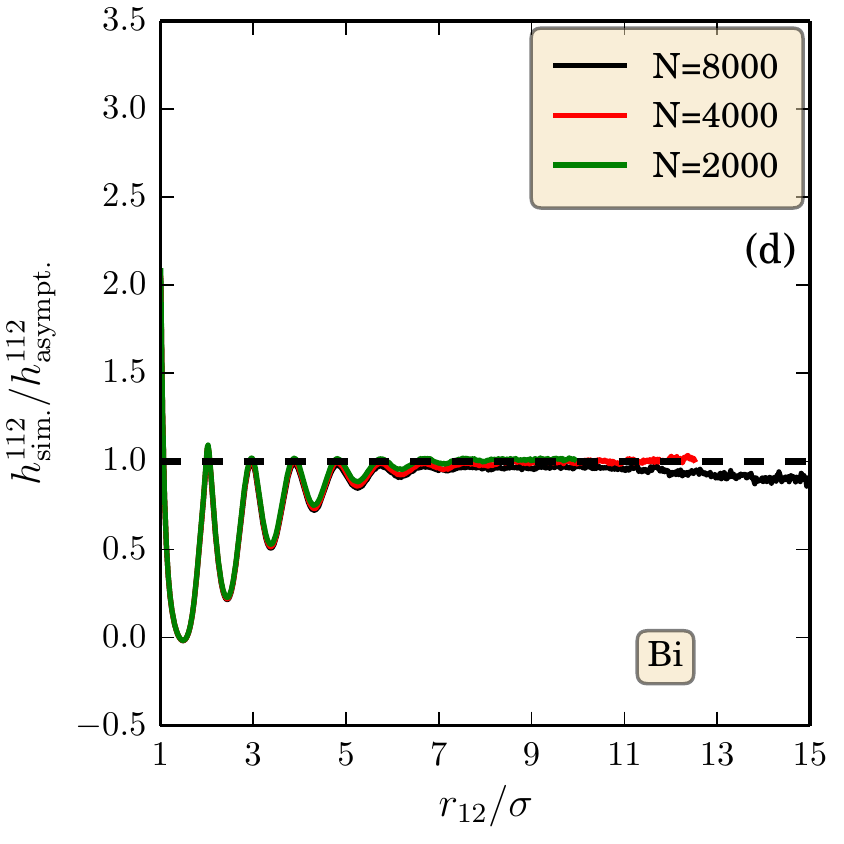}
\caption{Same as Fig.~\eqref{Fig5} with $\mu^{*2}=2.75$ }
\label{Fig6}
\end{figure}

As discussed in Sec.~\eqref{Asympto}  small differences should exist in the asymptotic regime which are
dictated by the geometry. Indeed, one finds the expected behavior~\eqref{assym}  as apparent in 
Figs.~(\ref{Fig5}) and (\ref{Fig6}).   The two  projections $h^{110}(r)$
and $h^{112(r)}$
tend towards their asymptotic predictions for distances $r_{12}>\zeta$ (\textit{i.e.}  $h^{mnl}_{\rm MC}/h^{mnl}_{\rm asympt.} \simeq 1$ 
as $r_{12} > \xi$), where $\xi \sim 7 \sigma$ for the two considered states.
Notice that the values of $h^{110}_{\rm MC.}/h^{110}_{\rm asympt.}$
at short separations diverges, as
$h^{110}_{\rm asympt.}\rightarrow 0$ when $R \rightarrow \infty$.

%%%%%%%%%%%%%%%%%%%%%%%%%%%%%%%%%%%%%%%%%%
%%%%%%%%%%%%%%%%%%%%%%%%%%%%%%%%%%%%%%%%%%

%%%%%%%%%%%%%%%%%%%%%%%%%%%%%%%%%%%%%%%%%%
%%%%%%%%%%%%%%%%%%%%%%%%%%%%%%%%%%%%%%%%%%

%%%%%%%%%%%%%%%%%%%%%%%%%%%%%%%%%%%%%%%%%%
%%%%%%%%%%%%%%%%%%%%%%%%%%%%%%%%%%%%%%%%%%

\section{Conclusion}
\label{Conclusion}
We have introduced a new method of simulation for dipolar liquids on the hypersphere $\mathcal{S}_3(0,R)$.
We have noted that, in this geometry, the electrostatics can be build in two different ways. Starting from
pseudo-charges we obtain mono-dipoles : the potential of such a dipole is a solution of Laplace-Beltrami
equation with a single singularity at the origin. Bi-dipoles are obtained from bi-charges and are more elaborated
since their electric potential, although also a solution of Laplace-Beltrami
equation, exhibits two singularities, one at the origin and the other at the antipodal point. 
A dipolar fluid 
can thus be represented as an assembly of mono-dipoles living in $\mathcal{S}_3(0,R)$ (the volume
of the system is then $2 \pi^2 R^3$) or a collection of  bi-dipoles living in the northern
hemisphere $\mathcal{S}_3(0,R)^+$ (the volume
of the system is now $ \pi^2 R^3$).
Of course these elaborated boundary conditions apply to ensure, in the absence of external
fields or walls, the homogeneity of the fluid : when a dipole leaves the
hemisphere $\mathcal{S}_3(0,R)^+$ at some point $M$ of the equator it reenters $\mathcal{S}_3(0,R)^+$ 
at the antipodal
point $\overline{M}$, bearing the same dipole. 
Of course mixture of bi-charges and bi-dipoles could be considered to simulate symmetric models
of electrolytes.

Since the Green's function of Laplace-Betrami is explicitely known both for mono- and bi-dipoles
the theory of the dielectric constant of the homogeneous fluid can be done in great details
in both cases, including
the derivation of the tail of the equilibrium pair correlation function induced by the curvature.
 Moreover we were able
to extract the contributions of these tail  to the Kirkwood's factor in the TL  and thus to
recover the well known Kirkwood's expression of the dielectric constant for
a fluid filling  the ordinary infinite Euclidian space.

We have reported MC data for simulations of two states of the fluid of dipolar hard spheres and
 performed systematic investigation of a novel potential on the hypersphere consisting of bi-dipoles. 
The great efficiency of the simulations on the hypersphere allows a drastic reduction
of the numerical uncertainties on the data. 
As far as the energy is concerned, the $1/N$ dependence on the MC data has been obtained for 
the largest systems yielding  precise estimates
of its thermodynamic limit,  with a relative precision of $\sim 10^{-4}$.
For both considered states, the TL  on the energy coincide
for mono and bi-dipoles in $\mathcal{S}_3$
within the error bars; they are also in good agreement
with the data obtained in cubico-periodic geometries together with the use of  Ewald potentials.
It seems that the TL limit is reached faster in the latter case, however the
relative precision on the Ewald potential (for a discussion see Ref.~\cite{Holm})
 is of the same order of magnitude than
the   precision that we obtained for the TL limit of the energy in our simulation in
$\mathcal{S}_3$, which should
temper the adepts of the former method.
The numerical  uncertainties on the dielectric constant preclude a similar study of its thermodynamic limit.
 As a remark : even at 
small system sizes (for instance $N=128$) one does not find errors greater than 2\% (compared to the TL) in 
energy and 20\% in the dielectric constant.

We have also checked the prediction of Sec.~\eqref{Asympto} on the behavior of the tails of the
angular correlation functions. The agreement between the theoretical predictions and the results
of simulations is quite excellent and gives insight on the (short) range of the 
dielectric tensor $\boldsymbol{\epsilon}(1,2)$.
Theoretical efforts are still needed to obtain an explicit expression of $\boldsymbol{\epsilon}(1,2)$
which should allow its complete calculation in a MC simulation.

It is difficult to assess the relative merits of mono or bi-dipoles. 
The convergence towards the TL
seems however  slightly, but marginally faster for bi-dipoles.
Clearly  both methods do not surpass the standard Ewald summation
techniques in size convergence.
However on the hypersphere the dipole-dipole
interactions can be computed, directly or with the help of tabulations, with an arbitrary precision by contrast
with the Ewald potential which allways involves  systematic numerical errors \cite{Holm}. In  cases where the precision on the pair potential
is crucial  the hypersphere technology should be preferred.

%%%%%%%%%%%%%%%%%%%%%%%%%%%%%%%%%%%%%%%%%%
%%%%%%%%%%%%%%%%%%%%%%%%%%%%%%%%%%%%%%%%%%

\begin{acknowledgments}
We thank D. Levesque and J.-J. Weis for providing us a MC code of dipolar
hard spheres with Ewald potentials that we used  for a part of the numerical
experiments on the state $(\rho^*=0.8, \mu^{*2}=2)$. J.-M. Caillol
acknowledges D. Levesque for his enlightened advises, interest and encouragements.
\end{acknowledgments}
%%%%%%%%%%%%%%%%%%%%%%%%%%%%%%%%%%%%%%%%%%
%%%%%%%%%%%%%%%%%%%%%%%%%%%%%%%%%%%%%%%%%%

\appendix*
\section{Some properties dipolar Green's functions in $\mathcal{S}_3$}
First we shall prove that, with  $M_1$ and $M_2$ being two points of the Northern hemisphere 
of the unit hypersphere $\mathcal{S}_3^+$, we have
\begin{equation}
\label{th1}
  \left[ \mathbf{G}_0 \circ   \mathbf{G}_0 \right] (1,2) =
 \ \int_{\mathcal{S}_3^+}d\, \Omega(3) \; \mathbf{G}_0(1,3) \cdot \mathbf{G}_0(3,2) = -\mathbf{G}_0(1,2) \; .
\end{equation}
We rewrite eq~\eqref{G0_b} as 
\begin{equation}
 \mathbf{G}_0(1,2) = \sum_{L \;, \boldsymbol{\alpha}} \,^{'} \mathbf{G}_0^{L, \boldsymbol{\alpha}} (1,2) \; ,
\end{equation}
with
\begin{equation}
  \mathbf{G}_0^{L, \boldsymbol{\alpha}} (1,2)= -\frac{2}{L(L+2)} 
\nabla_{\mathcal{S}_3}Y^*_{L,\boldsymbol{\alpha}}(\mathbf{z}_1) 
\nabla_{\mathcal{S}_3} Y_{L,\boldsymbol{\alpha}}(\mathbf{z}_2)  \; .
\end{equation}
Since for an  odd $L$, $Y_{L,\boldsymbol{\alpha}}(-\mathbf{z}) = - Y_{L,\boldsymbol{\alpha}}(\mathbf{z})$
one has
\begin{align}
\label{th2}
 \int_{\mathcal{S}_3^+}d\, \Omega(3) \; \mathbf{G}_0^{L, \boldsymbol{\alpha}}(1,3) 
 \cdot \mathbf{G}_0^{L^{'}, \boldsymbol{\alpha}^{'}} (3,2)& =
\frac{4}{L(L+2) L^{'}(L^{'}+2) }
 \nabla_{\mathcal{S}_3}Y^*_{L,\boldsymbol{\alpha}}(\mathbf{z}_1) 
  \nabla_{\mathcal{S}_3} Y_{L^{'},\boldsymbol{\alpha}^{'}}(\mathbf{z}_2)  \nonumber  \times \\
& \times \frac{1}{2} \int_{\mathcal{S}_3}d\, \Omega(3) 
  \nabla_{\mathcal{S}_3} Y_{L,\boldsymbol{\alpha}}(\mathbf{z}_3) \cdot
  \nabla_{\mathcal{S}_3}Y^*_{L^{'},\boldsymbol{\alpha^{'}}}(\mathbf{z}_3) \; .
\end{align}
The integral in Eq.~\eqref{th2} is computed by applying the Green-Beltrami identity~\eqref{Beltrami}
and the properties~\eqref{vp} and~\eqref{ortho} of the spherical harmonics giving us 

\begin{equation}
\label{th3}
 \int_{\mathcal{S}_3}d\, \Omega(\mathbf{z}) \;   \nabla_{\mathcal{S}_3}  Y_{L,\boldsymbol{\alpha}}(\mathbf{z}) \cdot
                        \nabla_{\mathcal{S}_3}Y^*_{L^{'},\boldsymbol{\alpha}^{'}}(\mathbf{z})       =
{L(L+2)} \delta_{L, L^{'}}     \delta_{\boldsymbol{\alpha},\boldsymbol{\alpha}^{'}}        \; .
\end{equation}
Inserting Eq.\eqref{th3} in Eq.\eqref{th2} readily yields the announced result~\eqref{th1}.

Our second result concern the integration of $\mathbf{G}_0(1,2)$ on a cone of axis $\mathbf{z}_1$
and aperture $0\leq \psi_0 \leq \pi/2$. We shall prove that
\begin{equation}
\label{int1}
 \int_{0 \leq \psi_{12}\leq \psi_0}d\, \Omega(\mathbf{z}_2) \; \mathbf{G}_0(1,2) =
 (-1+ \frac{2}{3} \cos \psi_0)
 \mathbf{U}_{\mathcal{S}_3}(\mathbf{z}_1) \; .
\end{equation}
To prove Eq.~\eqref{int1} one needs to take  some precaution because of the singularity of
$\mathbf{G}_0(1,2)$ at $\psi_{12}=\cos^{-1} ( \mathbf{z}_1 \cdot \mathbf{z}_2) \to 0$. We make use
of the decomposition~\eqref{decompo} to rewrite
\begin{equation}
\label{int2}
 \int_{0 \leq \psi_{12}\leq \psi_0}d\, \Omega(\mathbf{z}_2) \; \mathbf{G}_0(1,2) = -\frac{1}{3}
 \mathbf{U}_{\mathcal{S}_3}(\mathbf{z}_1)
+ \lim_{\delta \to 0}  \int_{ \delta\leq \psi_{12}\leq \psi_0}d\, \Omega(\mathbf{z}_2) \; \mathbf{G}_0(1,2) \, .
\end{equation}
The integral in the r.h.s. of Eq.~\eqref{int2} is computed by using spherical coordinates to reexpress the formula~\eqref{G0_b}
of the Green function and performing explicitely the integrals. A short computation gives us
\begin{equation}
\label{int3}
  \int_{ \delta\leq \psi_{12}\leq \psi_0}d\, \Omega(\mathbf{z}_2) \; \mathbf{G}_0(1,2)
= \frac{2}{3} (\cos \psi_0 - \cos \delta)  \mathbf{U}_{\mathcal{S}_3}(\mathbf{z}_1) \; ,
\end{equation}
with a well-behaved limit $\delta \to 0$.
Combining Eqs.~\eqref{int2} and~\eqref{int3} indeed yields the desired result~\eqref{int1}.

In this appendix we implicitely assumed that $R=1$. The reassessment of Eqs.~\eqref{th1} and~\eqref{int1} in
the case $R \neq 1$ is however trivial since the dipolar Green's function 
$\mathbf{G}_0(1,2)$ scales as $R^{-3}$ with the radius of the sphere. Clearly Eqs.~\eqref{th1} and~\eqref{int1}
remain valid for  $R \neq 1$ with the replacement $d \Omega(\mathbf{z}) \to d \tau(M)$ where
$d \tau(M)= R^3 d \Omega(\mathbf{z})$ is the infinitesimal volume element of the sphere $\mathcal{S}_3(O,R)$
of radius $R$.

%%%%%%%%%%%%%%%%%%%%%%%%%%%%%%%%%%%%%%%%%%%%%%%%%%%%%%%%%%%%%%
%%%%%%%%%%%%%%%%%%%%%%%%%%%%%%%%%%%%%%%%%%%%%%%%%%%%%%%%%%%%%%%

%%%%%%%%%%%%%%%%%%%%%%%%%%%%%%%%%%%%%%%%%%%%%%%%%%%%%%%
\end{document}